%% file: main.tex
\documentclass{article}

\usepackage[final]{neurips_2021}
\usepackage[utf8]{inputenc} % allow utf-8 input
\usepackage[T1]{fontenc}    % use 8-bit T1 fonts
\usepackage[hidelinks]{hyperref}
\usepackage{url}            % simple URL typesetting
\usepackage{booktabs}       % professional-quality tables
\usepackage{amsfonts}       % blackboard math symbols
\usepackage{nicefrac}       % compact symbols for 1/2, etc.
\usepackage{microtype}      % microtypography
\usepackage{xcolor}         % colors

\usepackage{graphicx}
\usepackage{comment}
\usepackage{amsmath}
\usepackage{amsthm}
\usepackage{amssymb}

\usepackage{float}
\usepackage{subfig}
\usepackage{algorithm}
\usepackage{algpseudocode}
\usepackage{tikz}
\usepackage{comment}
\usepackage{multirow}
\usepackage{comment}
\usepackage{cleveref}
\usetikzlibrary{arrows}

\graphicspath{{./figs/}}

\newtheorem{theorem}{Theorem}[section]
\newtheorem{lemma}[theorem]{Lemma}
\newtheorem{definition}[theorem]{Definition}

\definecolor{b2}{RGB}{51,153,255}
\definecolor{green}{RGB}{80,180,0}
\definecolor{yl}{RGB}{255,80,0}

\DeclareMathOperator{\R}{{\mathbb R}}

\newcommand{\wt}{\widetilde}

\newcommand{\diag}{\mathrm{diag}}

\title{Evaluating Gradient Inversion Attacks and Defenses in Federated Learning} 

\author{%
  Yangsibo Huang\\
  Princeton University\\
  Princeton, NJ 08540 \\
  \texttt{yangsibo@princeton.edu} \\
  
  \And
  Samyak Gupta\\
  Princeton University\\
  Princeton, NJ 08540 \\
  \texttt{samyakg@cs.princeton.edu} \\
  
  \And
  Zhao Song\\
  Adobe Research\\
  San Jose, CA 95110 \\
  \texttt{zsong@adobe.com} \\
  
  \And
  Kai Li\\
  Princeton University\\
  Princeton, NJ 08540 \\
  \texttt{li@cs.princeton.edu} \\
  
  \And
  Sanjeev Arora\\
  Princeton University\\
  Princeton, NJ 08540 \\
  \texttt{arora@cs.princeton.edu} \\
}

\begin{document}

\maketitle

\input{abstract}

\input{intro}

\input{related}

\input{exp}

\input{conclusion}

\newpage
\section*{Acknowledgments}
% \Yang{Kai and Sanjeev, please check if your funding sources are correct here:} 
This project is supported in part by 
% Kai's
Ma Huateng Foundation, Schmidt Foundation, 
% Sanjeev's
NSF, Simons Foundation, ONR and DARPA/SRC. 
Yangsibo Huang and Samyak Gupta are supported in part by the Princeton Graduate Fellowship.

We would like to thank Quanzheng Li, Xiaoxiao Li, Hongxu Yin and Aoxiao Zhong for helpful discussions, and members of Kai Li's and Sanjeev Arora's research groups for comments on early versions of the work. 

\bibliographystyle{plainnat}
\bibliography{ref}

\newpage

\input{app}

\end{document}

%% file: abstract.tex
\begin{abstract}
    Gradient inversion attack (or input recovery from gradient) is an emerging threat to the security and privacy preservation of Federated learning, whereby malicious eavesdroppers or participants in the protocol can recover (partially) the clients' private data. This paper evaluates existing attacks and defenses. We find that some attacks make strong assumptions about the setup. Relaxing such assumptions can substantially weaken these attacks. We then evaluate the benefits of three proposed defense mechanisms against gradient inversion attacks. We show the trade-offs of privacy leakage and data utility of these defense methods, and find that combining them in an appropriate manner makes the attack less effective, {\em even under the original strong assumptions}. We also estimate the computation cost of end-to-end recovery of a single image under each evaluated defense. Our findings suggest that the state-of-the-art attacks can currently be defended against with minor data utility loss, as summarized in a list of potential strategies. Our code is available at: \url{https://github.com/Princeton-SysML/GradAttack}.
\end{abstract}

%% file: intro.tex
\section{Introduction}

Federated learning~\citep{mcmahan2016communication,kairouz2019advances} is a framework that allows multiple clients in a distributed environment to collaboratively train a neural network model at a central server, without moving their data to the central server. 

At every training step, each client computes a model update ---i.e., gradient--- on its local data using the latest copy of the global model, and then sends the gradient to the central server. The server aggregates these updates (typically by averaging) to construct a global model, and then sends the new model parameters to all clients. 
%avoids the cost of raw data transferring, and also 
By allowing clients to participate in training without directly sharing their data, such protocols align better with data privacy regulations such as Health Insurance Portability and Accountability Act (HIPPA)~\citep{hippa}, California Consumer Privacy Act (CCPA)~\citep{ccpa}, and General Data Protection Regulation~\citep{gdpr}.

While  sharing gradients was thought to leak little information about the client's private data, recent papers~\citep{zhu2020deep, zhao2020idlg, geiping2020inverting, yin2021see}  developed a  ``gradient inversion attack'' by which an attacker eavesdropping on a client's communications with the server can begin to reconstruct the client's private data. 
The attacker can also be a malicious participant in the Federated Learning scheme, including a honest-but-curious server who wishes to reconstruct private data of clients, or a honest-but-curious client who wishes to reconstruct private data of other clients. These attacks have been shown to work with batch sizes only up to $100$ but even so they have created doubts about the level of privacy ensured in Federated Learning. The current paper seeks to evaluate the risks and suggest ways to minimize them. 

Several defenses against gradient inversion attacks have been proposed. These include perturbing gradients~\citep{zhu2020deep, wei2020framework} and using transformation for training data that clients can apply on the fly~\citep{zhang2017mixup, huang2020instahide}. More traditional cryptographic ideas including secure aggregation \citep{bonawitz2016practical} or homomorphic encryption \citep{phong18} for the gradients can also be used and presumably stop any eavesdropping attacks completely. They will not be studied here due to their special setups and overhead. 

We are not aware of a prior systematic evaluation of the level of risk arising from current attacks and the level of security provided by various defenses, as well as the trade-off (if any) between test accuracy, computation overhead, and privacy risks.

The paper makes two main contributions. First, we draw attention to two strong assumptions that a current gradient inversion attack~\citep{geiping2020inverting} implicitly makes. We show that by nullifying these assumptions, the performance of the attack drops significantly and can only work for low-resolution images. 
The findings are explored in  Section~\ref{sec:assumption} and already imply some more secure configurations in Federated Learning (Section~\ref{sec:discussion}).

Second, we summarize various defenses (Section~\ref{sec:defense}) and systematically evaluate (Section~\ref{sec:exp}) some of their performance of defending against a state-of-the-art gradient inversion attack, and present their data utility and privacy leakage trade-offs. We estimate the computation cost of end-to-end recovery of a single image under each evaluated defense. We also experimentally demonstrate the feasibility and effectiveness of combined defenses. Our findings are summarized as strategies to further improve Federated Learning's security against gradient inversion attacks (Section~\ref{sec:discussion}).

In Appendix~\ref{sec:app_theory}, we provide theoretical insights for mechanism of each evaluated defense.

%% file: related.tex
\section{Gradient Inversion Attacks}

Previous studies have shown the feasibility of recovering input from gradient (i.e. gradient inversion) for image classification tasks, by formulating it as an optimization problem: given a neural network with parameters $\theta$, and the gradient $ \nabla_\theta \mathcal{L}_{\theta}(x^*, y^*)$ computed with a private data batch $(x^*, y^*) \in \mathbb{R}^{b \times d} \times \mathbb{R}^{b}$ ($b, d$ being the batch size, image size), the attacker tries to recover $x \in \mathbb{R}^{b \times d}$, an approximation of $x^*$:

\begin{equation}
    \arg \min _{x}  \mathcal{L}_{\rm grad}(x; \theta, \nabla_{\theta} \mathcal{L}_{\theta}(x^{*}, y^{*})) + \alpha \mathcal{R}_{\rm aux}(x)
\end{equation}

The optimization goal consists of two parts: $\mathcal{L}_{\rm grad}(x; \theta, \nabla_{\theta} \mathcal{L}_{\theta}(x^{*}, y^{*}))$ enforces matching of the gradient of recovered batch $x$ with the provided gradients $\mathcal{L}_{\theta}(x^*, y^*)$, and $\mathcal{R}_{\rm aux}(x)$ regularizes the recovered image based on image prior(s). 

\citep{phong_17} brings theoretical insights on this task by proving that such reconstruction is possible with a single-layer neural network. 
\citep{zhu2020deep} is the first to show that accurate pixel-level reconstruction is practical for a maximum batch size of 8. Their formulation uses $\ell_2$-distance as $\mathcal{L}_{\rm grad}(\cdot, \cdot)$ but no regularization term $\mathcal{R}_{\rm aux}(x)$. The approach works for low-resolution CIFAR datasets~\citep{cifar10}, with simple neural networks with sigmoid activations, but cannot scale up to high-resolution images, or larger models with ReLU activations. A follow-up~\citep{zhao2020idlg} proposes a simple approach to extract the ground-truth labels from the gradient, which improves the attack but still cannot overcome its limitations.  

With a careful choice of $\mathcal{L}_{\rm grad}$ and $\mathcal{R}_{\rm aux}(x)$, \citep{geiping2020inverting} substantially improves the attack and succeeds in recovering a single ImageNet~\citep{deng2009imagenet} image from gradient: their approach uses cosine distance as $\mathcal{L}_{\rm grad}$, and the total variation as $\mathcal{R}_{\rm aux}(x)$. Their approach is able to reconstruct low-resolution images with a maximum batch size of 100 or a single high-resolution image. Based on~\citep{geiping2020inverting}, \citep{wei2020framework} analyzes how different configurations in the training may affect attack effectiveness. 

A more recent work~\citep{yin2021see} further improves the attack on high-resolution images, by introducing to $\mathcal{R}_{\rm aux}(x)$ a new image prior term based on batch normalization~\citep{ioffe2015batch} statistics,  and a regularization term which enforces consistency across multiple attack trials. 

An orthogonal line of work~\citep{zhu2020r} proposes to formulate the gradient inversion attack as a closed-form recursive procedure, instead of an optimization problem. However, their implementation can recover only low-resolution images under the setting where batch size = 1. 

\section{Strong Assumptions Made by SOTA Attacks}
\label{sec:assumption}

\subsection{The state-of-the-art attacks} 

Two recent attacks~\citep{geiping2020inverting,yin2021see} achieve best recovery results.  Our analysis focuses on the former as the implementation of the latter is not available at the time of writing this paper.
%Throughout this paper, we refer to as the state-of-the-art attack and focus on its evaluation.  
We plan to include the analysis for the latter attack in the final version of this paper if its implementation becomes available.

\citep{geiping2020inverting}'s attack optimizes the following objective function:
\begingroup
% \small
\begin{equation}
    \arg \min _{x}  1-\frac{\left\langle\nabla_{\theta} \mathcal{L}_{\theta}(x, y), \nabla_{\theta} \mathcal{L}_{\theta}\left(x^{*}, y^{*}\right)\right\rangle}{\left\|\nabla _ { \theta } \mathcal { L } _ { \theta } ( x , y ) \left|\left\|\mid \nabla_{\theta} \mathcal{L}_{\theta}\left(x^{*}, y^{*}\right)\right\|\right.\right.}+ \alpha_{\rm TV} \mathcal{R}_{\rm TV}(x)
\end{equation}
\endgroup

where $\langle \cdot, \cdot \rangle$ is the inner-product between vectors, and $\mathcal{R}_{\rm TV}(\cdot)$ is the total variation of images.

We notice that Geiping et al. has made two strong assumptions (Section~\ref{sec:assumption_explain}). Changing setups to invalidate those assumptions will substantially weaken the attacks (Section~\ref{sec:assumption_invalidate}). We also summarize whether other attacks have made similar assumptions in Table~\ref{tab:assumption_summary}.

\subsection{Strong assumptions} 
\label{sec:assumption_explain}

We find that previous gradient inversion attacks have made different assumptions about whether the attacker knows Batch normalization statistics or private labels, as shown in  Table~\ref{sec:assumption_explain}. 
Note that \citep{geiping2020inverting}'s attack makes both strong assumptions.

\paragraph{Assumption 1: Knowing BatchNorm statistics.} Batch normalization (BatchNorm) \citep{ioffe2015batch} is a technique for training neural networks that normalizes the inputs to a layer for every mini-batch. It behaves differently during training and evaluation. Assume the model has $L$ batch normalization layers. Given $x^*$, a batch of input images, we use $x^*_l$ to denote the input features to the $l$-th BatchNorm layer, where $l \in [L]$. During training, the $l$-th BatchNorm layer normalizes $x^*_l$ based on the batch's mean ${\rm mean}(x^*_l)$ and variance ${\rm var}(x^*_l)$, and keeps a running estimate of mean and variance of all training data points, denoted by $\mu_l$ and $\sigma^2_l$. 
During inference, $\{\mu_l\}_{l=1}^L$ and $\{\sigma^2_l\}_{l=1}^L$ are used to normalize test images. In the following descriptions, we leave out $\{\cdot\}_{l=1}^L$ for simplicity (i.e. use $\mu, \sigma^2$ to denote $\{\mu_l\}_{l=1}^L, \{\sigma^2_l\}_{l=1}^L$, and ${\rm mean}(x^*)$, ${\rm var}(x^*)$ to denote $\{{\rm mean}(x^*_l)\}_{l=1}^L$, $\{{\rm var}(x^*_l)\}_{l=1}^L$).

We notice that \citep{geiping2020inverting}'s implementation\footnote{The official implementation of \citep{geiping2020inverting}: \href{https://github.com/JonasGeiping/invertinggradients}{https://github.com/JonasGeiping/invertinggradients}.} assumes that BatchNorm statistics of the  private batch, i.e., ${\rm mean}(x^*)$, ${\rm var}(x^*)$, are jointly provided with the gradient. Knowing BatchNorm statistics would enable the  attacker to apply the same batch normalization used by the private batch on his recovered batch, to achieve a better reconstruction.  This implicitly increases the power of the attacker, as sharing private BatchNorm statistics are not necessary in Federated learning~\citep{andreux2020siloed, li2021fedbn}.

Note that this assumption may be \textit{realistic} in some settings:  1) the neural network is shallow, thus does not require using BatchNorm layers, or 2) the neural network is deep, but adapts approaches that normalize batch inputs with a fixed mean and variance (as alternative to BatchNorm), e.g. Fixup initialization \citep{zhang2019fixup}.

\paragraph{Assumption 2: Knowing or able to infer private labels.} Private labels are not intended to be shared in Federated learning, but knowing them would improve the attack. \citep{zhao2020idlg} finds that label information of a {\em single} private image can be inferred from the gradient (see Section~\ref{sec:assumption_invalidate} for details). Based on this, \citep{geiping2020inverting} assumes the attacker knows private labels (see remark at the end of Section 4 in their paper).  However, this assumption may not hold true when multiple images in a batch share the same label, as we will show in the next section. 

\begin{table}[t]
    \small
    \centering
    \setlength{\tabcolsep}{3.8pt}
    \begin{tabular}{|l|cccc|}
    \toprule
        {\bf Assumptions} & \citep{zhu2020deep} & \citep{zhao2020idlg} & \citep{geiping2020inverting} & \citep{yin2021see}\\
    \midrule
    {\bf Knowing BN statistics} & N/A$^\dagger$ & N/A$^\dagger$ & Yes & Yes$^*$\\
    {\bf Knowing private labels } &  No & No$^\ddagger$ & Yes & No$^\ddagger$\\
    \bottomrule
    \end{tabular}
    \vspace{2mm}
    \caption{Assumptions of gradient inversion attacks.
    {%\footnotesize
    $^\dagger$: its evaluation uses a simple model without a BatchNorm layer;  $^\ddagger$: it proposes a method to infer private labels, which works when images in a batch have unique labels (see Section~\ref{sec:assumption_invalidate}); $^*$: although the paper discusses a setting where BatchNorm statistics are unknown, its main results assume knowing BatchNorm statistics. 
    }
    }
    \label{tab:assumption_summary}
    \vspace{-3mm}
\end{table}

\subsection{Re-evaluation under relaxed assumptions} 
\label{sec:assumption_invalidate}

We re-evaluate the performance of the gradient inversion attack in settings where two assumptions above are relaxed. For each relaxation, we re-design the attack (if needed) based on the knowledge that the attacker has.

\paragraph{Relaxation 1: Not knowing BatchNorm statistics.} We refer to the previous threat model as $\rm BN_{exact}$, where the attacker knows exact BatchNorm statistics of the private batch. We consider a more realistic threat model where these statistics are not exposed, and re-design the attack based on it.

\textit{Threat model.} In each training step, the client normalizes its private batch $x^*$ using the batch's mean ${\rm mean}(x^*)$ and variance ${\rm var}(x^*)$, keeps the running estimate of mean and variance \textit{locally} as in \citep{li2021fedbn}, and shares the gradient. The client releases the final aggregated mean $\mu$, and aggregated variance $\sigma^2$ of all training data points at the end of training. Same as before, the attacker has access to the model and the gradient during training. %He uses $\mu$ and $\sigma^2$ to help him launch the gradient inversion attack.

\begin{figure}[H]
% \vspace{-3mm}
    \centering
    \subfloat[Original]{\includegraphics[width=0.123\textwidth]{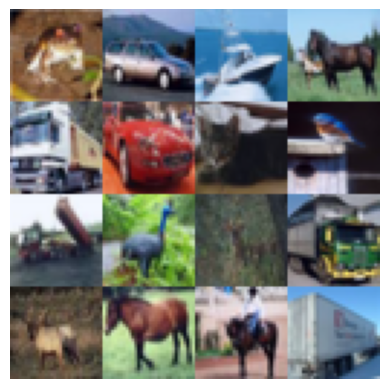}}
    \subfloat[$\rm BN_{exact}$]{\includegraphics[width=0.123\textwidth]{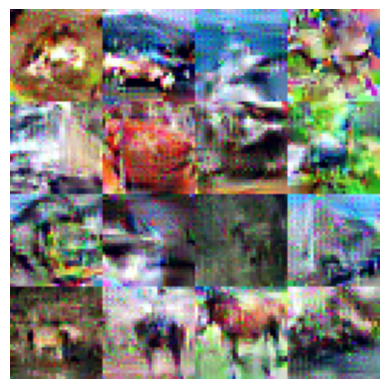}}
    \subfloat[$\rm BN_{infer}$]{\includegraphics[width=0.123\textwidth]{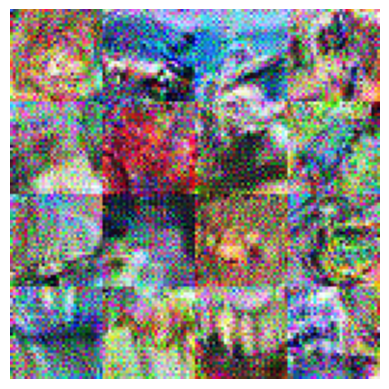}}
    \subfloat[$\rm BN_{proxy}$]{\includegraphics[width=0.123\textwidth ]{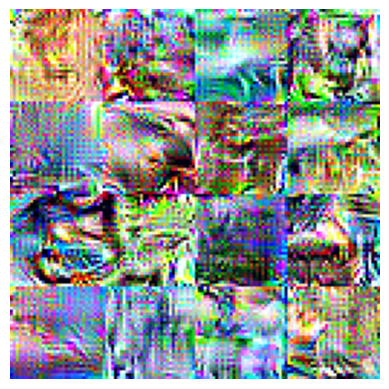}}
    \hspace{1mm}
    \subfloat[Original]{\includegraphics[width=0.123\textwidth]{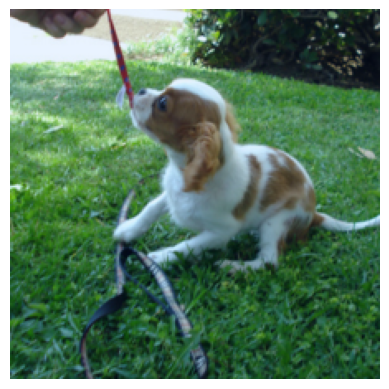}}
    \subfloat[$\rm BN_{exact}$]{\includegraphics[width=0.123\textwidth]{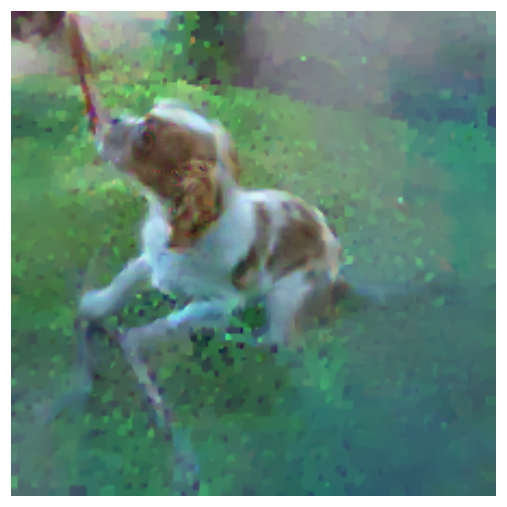}}
    \subfloat[$\rm BN_{infer}$]{\includegraphics[width=0.123\textwidth]{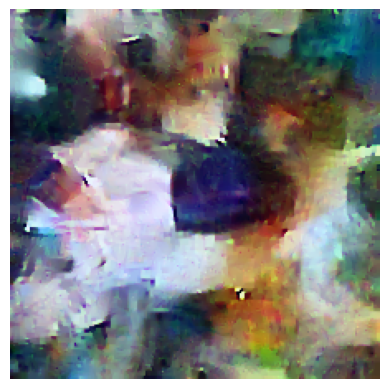}}
    \subfloat[$\rm BN_{proxy}$]{\includegraphics[width=0.123\textwidth ]{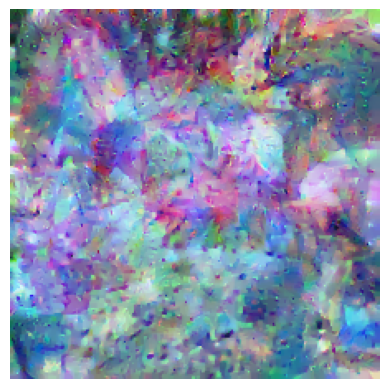}}
    \caption{ Attacking a batch of 16 low-resolution images from CIFAR-10 (a-d) and a single high-resolution image from ImageNet (e-h) with different knowledge of BatchNorm statistics. 
    Attack is weakened when BatchNorm statistics are not available (c, d versus b, and g, h versus f). See Appendix~\ref{sec:app_exp} for more examples and quantitative results.}
    \label{fig:BN_var}
\end{figure}

\textit{Re-design A: $\rm BN_{proxy}$, attacker naively uses $\mu$ and $\sigma^2$.} A simple idea is that the attacker uses $(\mu, \sigma^2)$ as the proxy for $({\rm mean}(x^*), {\rm var}(x^*))$, and uses them to normalize $x$, his guesses of the private batch. Other operations of the gradient inversion attack remain the same as before. However, Figure~\ref{fig:BN_var}.d and~\ref{fig:BN_var}.h show poor-quality reconstruction with this re-design.

\textit{Re-design B: $\rm BN_{infer}$, attacker infers $({\rm mean}(x^*), {\rm var}(x^*))$ based on $(\mu, \sigma^2)$.} %The failure of $\rm BN_{proxy}$ is mainly due to its non-adaptive way of using $(\mu, \sigma^2)$. 
A more reasonable attacker will try to infer $({\rm mean}(x^*), {\rm var}(x^*))$ while updating $x$, his guesses of the private batch, and uses $({\rm mean}(x), {\rm var}(x))$ to normalize the batch. In this case, $(\mu, \sigma^2)$ could be used as a prior of BatchNorm statistics to regularize the recovery, as suggested in \citep{yin2021see}:

\begingroup
% \small
\begin{equation}
\label{eq:objective}
    \arg \min _{x}  1-\frac{\left\langle\nabla_{\theta} \mathcal{L}_{\theta}(x, y), \nabla_{\theta} \mathcal{L}_{\theta}\left(x^{*}, y^{*}\right)\right\rangle}{\left\|\nabla _ { \theta } \mathcal { L } _ { \theta } ( x , y ) \left|\left\|\mid \nabla_{\theta} \mathcal{L}_{\theta}\left(x^{*}, y^{*}\right)\right\|\right.\right.}+ \alpha_{\rm TV} \mathcal{R}_{\rm TV}(x) + \alpha_{\rm BN} \mathcal{R}_{\rm BN}(x)
\end{equation}
% \vspace{-4mm}
\endgroup

where $ \mathcal{R}_{\mathrm{BN}}(x)=  \sum_{l} \| {\rm mean}(x_l)- \mu_l \|_{2} + \sum_{l} \| {\rm var}(x_l)- \sigma_l^2 \|_{2}$.

We tune $\alpha_{\rm BN}$ and present the best result in Figure~\ref{fig:BN_var}.c and~\ref{fig:BN_var}.g (see results of different $\alpha_{\rm BN}$'s in Appendix~\ref{sec:app_exp}). As shown, for a batch of low-resolution images, $\rm BN_{infer}$ gives a much better reconstruction result than $\rm BN_{proxy}$, but still cannot recover some details of the private batch when compared with $\rm BN_{exact}$. The result for a single high-resolution image is worse: the attacker fails to return a recognizable reconstruction with $\rm BN_{infer}$. This suggests not having access to BatchNorm statistics of the private batch already weakens the state-of-the-art gradient inversion attack.

% \captionsetup[figure]{font=small}

\begin{figure}[t]
    \centering
    \subfloat[Distribution of labels in a batch]{
        \includegraphics[width=.35\textwidth]{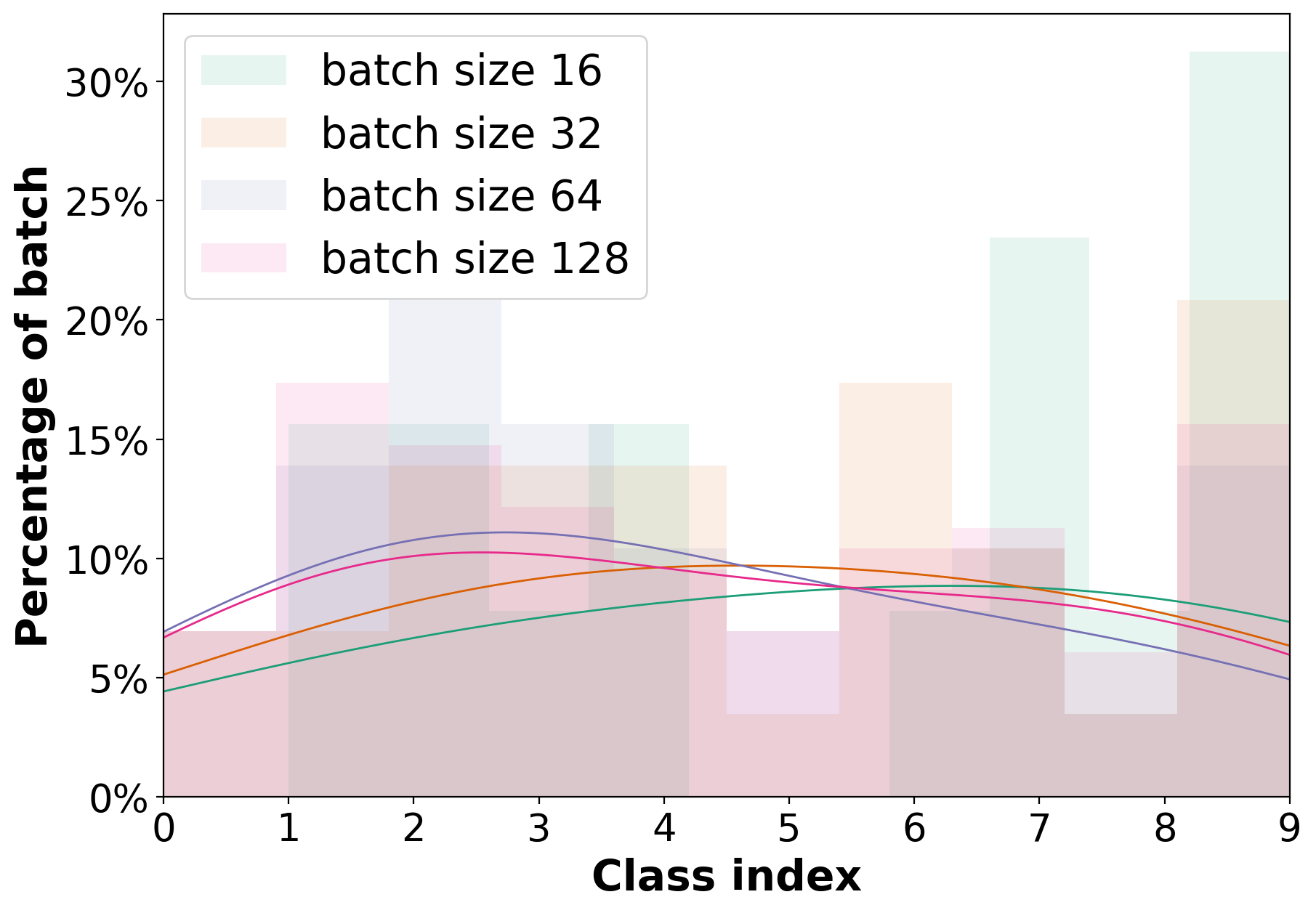}
        \label{fig:batch_label_dist}
    }
    \subfloat[Reconstructions with and without private labels]{
    \includegraphics[width=.6\textwidth]{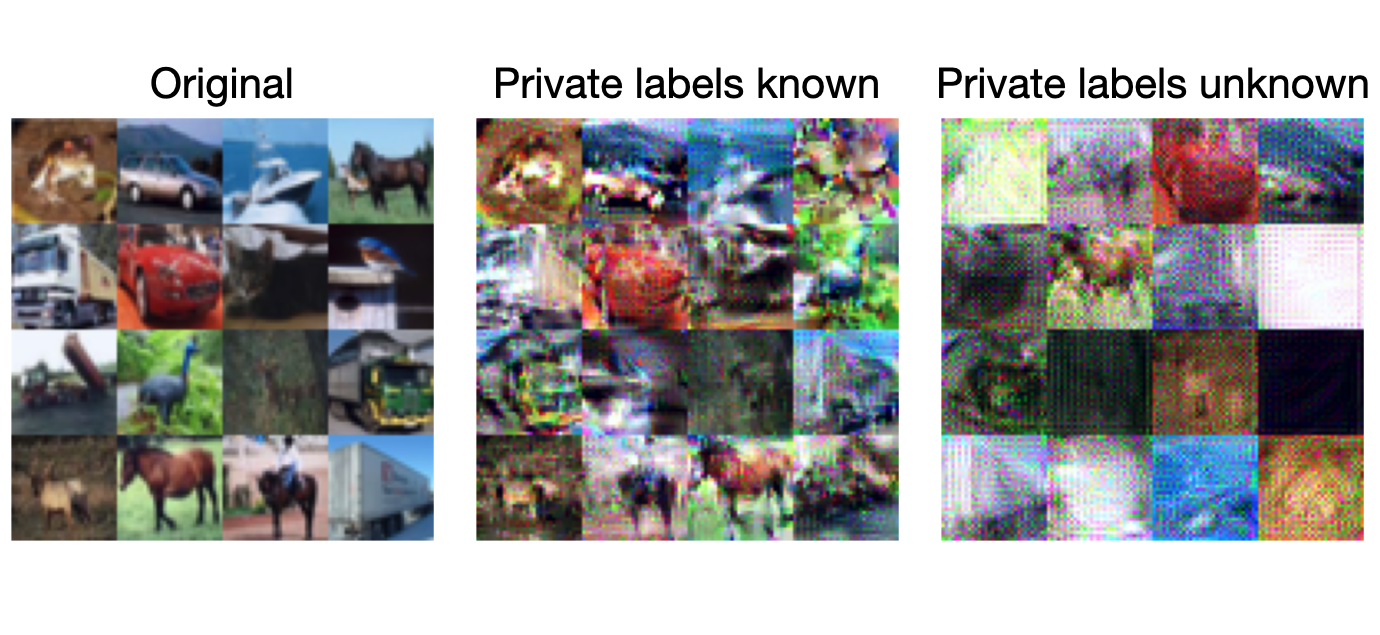}
    \label{fig:reconstructed_labels}
    }
    \caption{ Attack is weakened when private labels are not available. (a) shows that for CIFAR-10, when the batch size is large, many images in the batch belong to the same class, which essentially weakens label restoration~\citep{zhao2020idlg, yin2021see}. (b) visualizes a reconstructed batch of 16 images with and without private labels known. The quality of the reconstruction drops without knowledge of private labels. 
    }
    \vspace{-3mm}
\end{figure}

\paragraph{Relaxation 2: Not knowing private labels.}  \citep{zhao2020idlg} notes that label information of a {\em single} private image can be computed analytically from the gradients of the layer immediately before the output layer. \citep{yin2021see} further extends this method to support recovery of labels for a batch of images. However, if multiple images in the private batch belong to a same label, neither approach can tell how many images belong to that label, let alone which subset of images belong to that label. Figure~\ref{fig:batch_label_dist} demonstrates that with CIFAR-10, for batches of various sizes it is possible for many of the training samples to be have the same label, and the distribution of labels is not uniform - and hence, inferring labels becomes harder and the attack would be weakened. In Figure~\ref{fig:reconstructed_labels} we evaluate the worst-case for an attacker in this setting by comparing recoveries where the batch labels are simultaneously reconstructed alongside the training samples.

\section{Defenses Against the Gradient Inversion Attack}
\label{sec:defense}

Several defense ideas have been proposed to mitigate the risks of gradient inversion.

\subsection{Encrypt gradients}

Cryptography-based approaches encrypt gradient to prevent gradient inversion. \citep{bonawitz2016practical} presents a secure aggregation protocol for Federated learning by computing sum of gradient vectors based on secret sharing \citep{shamir1979share}. \citep{phong18} proposes using homomorphic encryption to encrypt the gradients before sending. These approaches require special setup and can be costly to implement.

Moreover, with secure aggregation protocol, an honest-but-curious server can still launch the gradient inversion attack on the summed gradient vector. Similarly, an honest-but-curious client can launch the gradient inversion attack on the model returned by the server to reconstruct other clients' private data, even with homomorphic encryption.

As alternatives, two other types of defensive mechanisms have been proposed to mitigate the risks of attacks on \textit{plain-text} gradient.

\subsection{Perturbing gradients}

\paragraph{Gradient pruning.} When proposing the first practical gradient inversion attack, \citep{zhu2020deep} also suggests a defense by setting gradients of small magnitudes to zero (i.e. gradient pruning). Based on their attack, they demonstrate that pruning more than $70\%$ of the gradients would make the recovered images no longer visually recognizable. However, the suggested prune ratio is determined based on weaker attacks, and may not remain safe against the state-of-the-art attack.

\paragraph{Adding noise to gradient.} Motivated by DPSGD \citep{abadi2016deep} which adds noise to gradients to achieve differential privacy~\citep{d09, dr14}, \citep{zhu2020deep, wei2020framework} also suggests defending by adding Gaussian or Laplacian noise to gradient. They show that a successful defense requires adding high noise level such that its accuracy drops by more than $30\%$ with CIFAR-10 tasks. Recent works \citep{papernot2020making, tramer2020differentially} suggest using better pre-training techniques and a large batch size (e.g. $4,096$) to achieve a better accuracy for DPSGD training.

Since most DPSGD implementations for natural image classification tasks~\citep{abadi2016deep, papernot2020making, tramer2020differentially} use a pre-training and fine-tuning pipeline, it is hard to fairly compare with other defense methods that can directly apply when training the model from scratch. Thus, we leave the comparison with DPSGD to future work.

\subsection{Weak encryption of inputs (i.e. encoding inputs)}

\paragraph{MixUp.} MixUp data augmentation \citep{zhang2017mixup} trains neural networks on composite images created via linear combination of image pairs. It has been shown to improve the generalization of the neural network and stabilizes the training. Recent work also suggests that MixUp increases the model's robustness to adversarial examples~\citep{pang2019mixup, lamb2019interpolated}. 

\paragraph{InstaHide.} Inspired by MixUp, \citep{huang2020instahide} proposes InstaHide as a light-weight instance-encoding scheme for private distributed learning. To encode an image $x \in {\mathbb R}^d$ from a private dataset, InstaHide first picks $k-1$ other images $s_2, s_3, \ldots, s_k$ from that private dataset, or a large public dataset, and $k$ random nonnegative coefficients $\{\lambda_i\}_{i=1}^k$ that sum to $1$, and creates a composite image $\lambda_1 x + \sum_{i=2}^k \lambda_i s_i$ ($k$ is typically small, e.g., $4$). A composite label is also created using the same set of coefficients.\footnote{Only the labels of examples from the private dataset will get combined. See \citep{huang2020instahide} for details.} Then it adds another layer of security: pick a random sign-flipping pattern $\sigma \in \{-1, 1\}^d$ and output the encryption $\tilde{x} = \sigma\circ (\lambda_1 x + \sum_{i=2}^k \lambda_i s_i)$, where $\circ$ is coordinate-wise multiplication of vectors.
The neural network is then trained on encoded images, which look like random pixel vectors to the human eye and yet lead to  good classification accuracy ($<6\%$ accuracy loss on CIFAR-10, CIFAR-100, and ImageNet).

Recently, \citep{carlini2020attack} gives an attack to recover private images of a small dataset, when the InstaHide encodings are revealed to the attacker (not in a Federated learning setting). Their first step is to train a neural network on a public dataset for similarity annotation, 
to infer whether a pair of InstaHide encodings contain the same private image. With the inferred similarities of all pairs of encodings, the attacker then runs a combinatorial algorithm (cubic time in size of private dataset) to cluster all encodings based on their original private images, and finally uses a regression algorithm (with the help of composite labels) to recover the private images. 

Neither \citep{huang2020instahide} or \citep{carlini2020attack} has evaluated their defense or attack in the Federated learning setting, where the attacker observes gradients of the encoded images instead of original encoded images. This necessitates the systematic evaluation in our next section. 

%% file: exp.tex
\section{Evaluation of defenses}
\label{sec:exp}

The main goal of our experiments is to understand the trade-offs between data utility (accuracy) and securely defending the state-of-the-art gradient inversion attack even in its strongest setting, \textit{without} any relaxation of its implicit assumptions. Specifically, we grant the attacker the knowledge of 1) BatchNorm statistics of the private batch, and 2) labels of the private batch. 

We vary key parameters for each defense, and evaluate their performance in terms of the test accuracy, computation overhead, and privacy risks (Section~\ref{sec:exp_single}). We then investigate the feasibility of combining defenses (Section~\ref{sec:exp_combine}). We also estimate the computation cost of end-to-end recovery of a single image under evaluated defenses (Section~\ref{sec:exp_estimate}).

As the feasibility of the state-of-the-art attack~\citep{geiping2020inverting} on a batch of high-resolution images remain elusive when its implicit assumptions no longer hold (see Figure~\ref{fig:BN_var}), we focus on the evaluation with low-resolution in trying to understand whether current attacks can be mitigated.

\subsection{Experimental setup} 
\label{sec:exp_setup}

\paragraph{Key parameters of defenses.} We evaluate following defenses on CIFAR-10 dataset \citep{cifar10} with ResNet-18 architecture \citep{he2015resnet}. 

\begin{itemize}
  \item {\bf GradPrune} (gradient pruning): gradient pruning set gradients of small magnitudes to zero. We vary the pruning ratio $p$ in $\{0.5, 0.7, 0.9, 0.95, 0.99, 0.999\}$. 
%   \vspace{-1mm}
  \item {\bf MixUp}: MixUp encodes a private image by linearly combining it with $k-1$ other images from the training set. Following~\citep{huang2020instahide}, we vary $k$ in $\{4, 6\}$, and set the upper bound of a single coefficient to $0.65$ (coefficients sum to $1$). 
%   \vspace{-1mm}
  \item {\bf Intra-InstaHide}: InstaHide ~\citep{huang2020instahide} proposes two versions: Inter-InstaHide and Intra-InstaHide. The only difference is that at the mixup step, Inter-Instahide mixes up an image with images from a public dataset, whereas Intra-InstaHide only mixes with private images. Both versions apply a random sign flipping pattern on each mixed image.
  We evaluate Intra-InstaHide in our experiments, which is a weaker version of InstaHide. Similar to the evaluation of MixUp, we vary $k$ in $\{4, 6\}$, and set the upper bound of a single coefficient to $0.65$. Note that InstaHide flips signs of pixels in the image, which destroys the total variation prior. However, the absolute value of adjacent pixels should still be close. Therefore, for the InstaHide defense, we apply the total variation regularizer on $|x|$, i.e. taking absolute value of each pixel in the reconstruction.
\end{itemize}

We train the ResNet-18 architecture on CIFAR-10 using different defenses, and launch the attack. We provide more details of the experiments in Appendix~\ref{sec:app_exp}. 

\paragraph{The attack.} We use a subset of 50 CIFAR-10 images to evaluate the attack performance. Note that attacking MixUp and InstaHide involves another step to decode private images from the encoded images. We apply \citep{carlini2020attack}'s attack here as the decode step, where the attacker needs to eavesdrop $T$ epochs of training, instead of a single training step. We set $T=20$ in our evaluation. We also grant the attacker the strongest power for the decode step to evaluate the upper bound of privacy leakage. Given a MixUp or Intra-InstaHide image which encodes $k$ private images, we assume the attacker knows:
\begin{enumerate}
% \vspace{-2mm}
  \item The indices of $k$ images in the private dataset. In a realistic scenario, the attacker of \citep{carlini2020attack} would need to train a neural network to detect similarity of encodings, and run a combinatorial algorithm to solve {\em an approximation of} this mapping.
%   \vspace{-2mm}
  \item The mixing coefficients for each of the $k$ private image. In real Federated learning, this information is \textit{not available}.
% \vspace{-2mm}
\end{enumerate}

\paragraph{Hyper-parameters of the attack.} The attack minimize the objective function given in Eq.\ref{eq:objective}. 
We search $\alpha_{\rm TV}$ in $\{0, 0.001, 0.005, 0.01, 0.05, 0.1, 0.5\}$ for all defenses, and apply the best choice for each defense: $0.05$ for GradPrune, $0.1$ for MixUp, and $0.01$ for Intra-InstaHide. We apply $\alpha_{\rm BN}=0.001$ for all defenses after searching it in $\{0, 0.0005, 0.001, 0.01, 0.05, 0.01\}$. We optimize the attack for $10,000$ iterations using Adam~\citep{kingma2014adam}, with initial learning rate $0.1$. We decay the learning rate by a factor of $0.1$ at $3/8, 5/8, 7/8$ of the optimization. 

\paragraph{Batch size of the attack.} \citep{zhu2020deep, geiping2020inverting} have shown that a small batch size is important for the success of the attack. We intentionally evaluate the attack with three small batch sizes to test the upper bound of privacy leakage, including the minimum (and unrealistic) batch size 1, and two small but realistic batch sizes, 16 and 32.

\paragraph{Metrics for reconstruction quality.} We visualize reconstructions obtained under different defenses. Following~\citep{yin2021see}, we also use the learned perceptual image patch similarity (LPIPS) score~\citep{zhang2018unreasonable} to measure mismatch between reconstruction and original images: higher values suggest more mismatch (less privacy leakage).

\begin{figure}[ht]
  \centering
  \includegraphics[width=\linewidth]{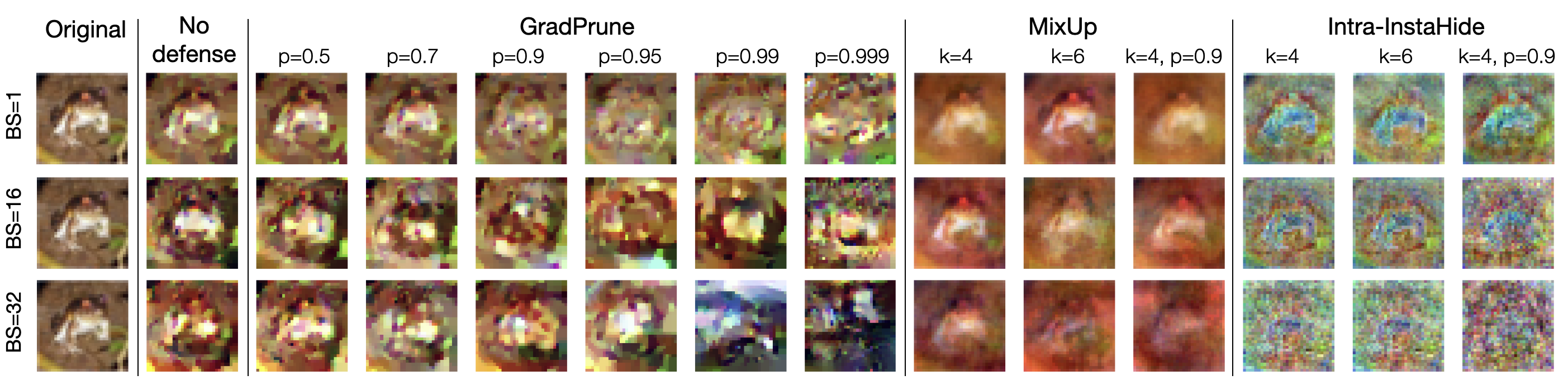}
  \caption{Reconstruction results under different defenses with batch size being 1, 16 and 32. When batch size is 32, combining gradient pruning and Intra-InstaHide makes the reconstruction almost unrecognizable (the last column). See Figure~\ref{fig:vis_recon_full} in Appendix~\ref{sec:app_exp} for the full version.}
  \label{fig:vis_recon}
%   \vspace{2mm}
\end{figure}

\newcommand{\best}[1]{\textcolor{green}{\textbf{#1}}}
% \captionsetup[table]{font=small}
\begin{table}[t] 
  \scriptsize
  \setlength{\tabcolsep}{2.8pt}
  \renewcommand{\arraystretch}{0.95}
  \begin{tabular}{|l|c|c|c|c|c|c|c|c|c|c|c|c|c|c|c|c|}
  \toprule
   &  \multirow{2}{*}{\bf None} & \multicolumn{6}{c|}{\multirow{2}{*}{\bf GradPrune ($p$)}} & \multicolumn{2}{c|}{\multirow{2}{*}{\bf MixUp ($k$)}} & \multicolumn{2}{c|}{\multirow{2}{*}{\bf Intra-InstaHide ($k$)}} & \multicolumn{2}{c|}{\bf GradPrune ($p=0.9$)}\\
   & & \multicolumn{6}{c|}{} & \multicolumn{2}{c|}{} & \multicolumn{2}{c|}{} & {\bf  + MixUp } & {\bf  + Intra-InstaHide}\\
  \midrule
   {\bf Parameter}  & - & 0.5 & 0.7 & 0.9 & 0.95 & 0.99 & 0.999 & 4 & 6 & 4 & 6 & $k=4$ & $k=4$ \\
   \midrule
   {\bf Test Acc.} & 93.37 & 93.19 & 93.01 & 90.57 & 89.92 & 88.61 & 83.58 &  92.31 & 90.41 & 90.04 & 88.20 & 91.37 & 86.10 \\
   \midrule
   {\bf Time (train)} & $1\times$ & \multicolumn{6}{c|}{$1.04\times$} & \multicolumn{2}{c|}{$1.06\times$} & \multicolumn{2}{c|}{$1.06\times$} & \multicolumn{2}{c|}{$1.10\times$} \\
   \midrule
   \multicolumn{14}{|c|}{\bf Attack batch size $= 1$} \\
   \midrule
   {\bf Avg. LPIPS $\downarrow$}  & 0.19  & 0.19  & 0.22  & 0.35  & 0.42  & 0.52  & 0.52  & 0.34  & 0.46  & 0.58  & \best{0.61}  & 0.41  & 0.60\\
   {\bf Best LPIPS $\downarrow$}  & 0.02  & 0.02  & 0.05  & 0.14  & 0.22  & 0.32  & 0.36  & 0.12  & 0.25  & 0.41  & 0.42  & 0.21  & \best{0.43}\\
   {(LPIPS std.)}                 & 0.16  & 0.17  & 0.16  & 0.13  & 0.11  & 0.08  & 0.06  & 0.08  & 0.07  & 0.06  & 0.09  & 0.07  & 0.09\\
   \midrule
   \multicolumn{14}{|c|}{\bf Attack batch size $= 16$} \\
   \midrule
   {\bf Avg. LPIPS $\downarrow$}  & 0.45  & 0.46  & 0.47  & 0.51  & 0.55  & 0.58  & 0.61    & 0.34  & 0.31  & 0.62  & 0.63  & 0.46  & \best{0.68}\\
   {\bf Best LPIPS $\downarrow$}  & 0.18  & 0.19  & 0.19  & 0.31  & 0.43  & 0.47  & 0.51    & 0.11  & 0.13  & 0.41  & 0.44  & 0.22  & \best{0.54}\\
   {(LPIPS std.)}                 & 0.12  & 0.12  & 0.11  & 0.07  & 0.05  & 0.04  & 0.03    & 0.09  & 0.09  & 0.08  & 0.08  & 0.10  & 0.07\\
   \midrule
   \multicolumn{14}{|c|}{\bf Attack batch size $= 32$} \\
   \midrule
   {\bf Avg. LPIPS $\downarrow$}  & 0.45  & 0.46  & 0.48  & 0.52  & 0.54  & 0.58  & 0.63         & 0.50  & 0.49  & 0.69  & 0.69  & 0.62  & \best{0.73}\\
   {\bf Best LPIPS $\downarrow$}  & 0.18  & 0.18  & 0.22  & 0.31  & 0.43  & 0.48  & 0.54         & 0.31  & 0.28  & 0.56  & 0.56  & 0.37  & \best{0.65}\\
   {(LPIPS std.)}                 & 0.11  & 0.11  & 0.09  & 0.07  & 0.05  & 0.04  & 0.04         & 0.10  & 0.10  & 0.06  & 0.07  & 0.10  & 0.05\\
  \bottomrule
  \end{tabular}
  \subfloat{\includegraphics[width=0.45\linewidth]{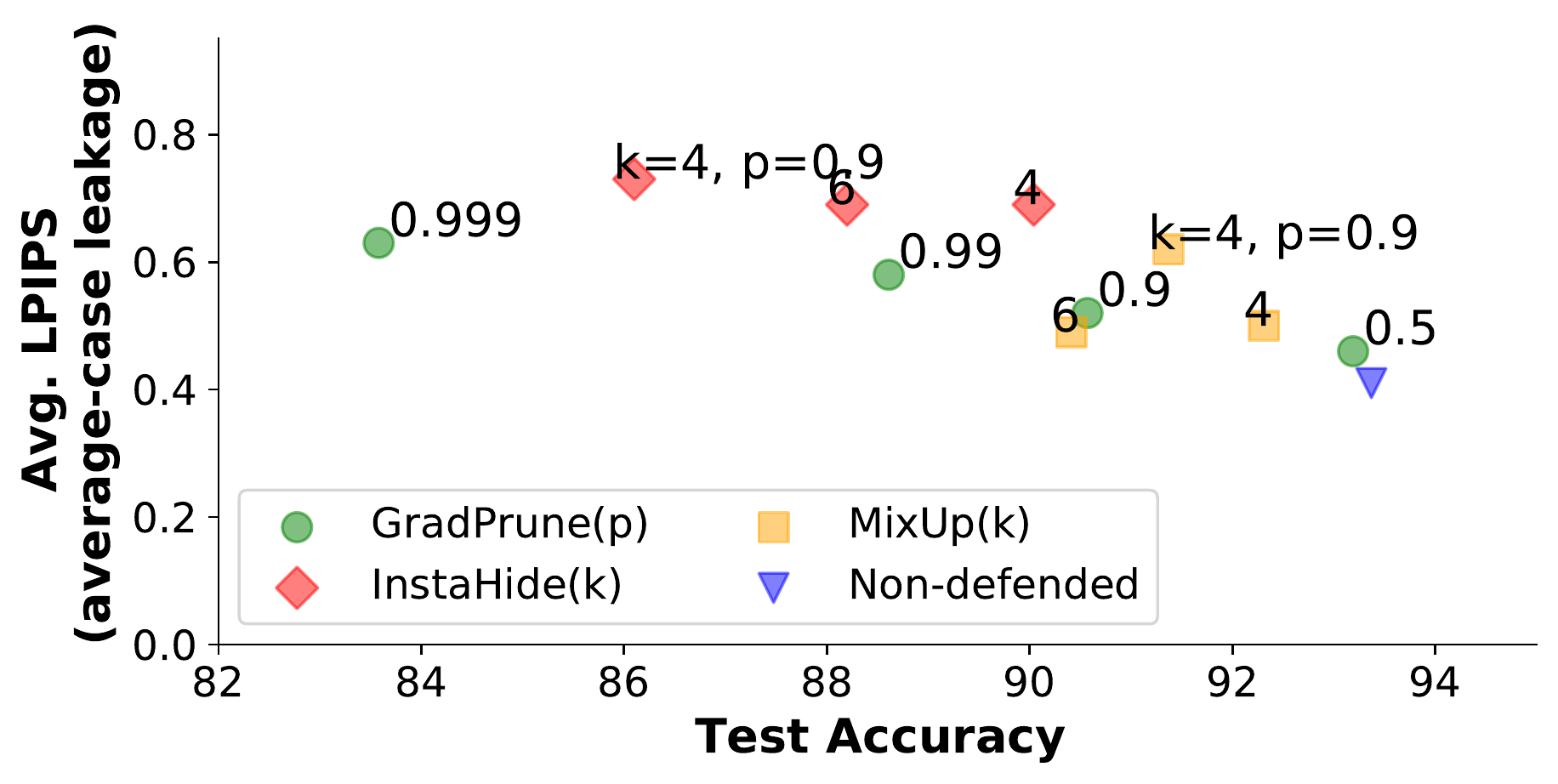}}
  \hspace{7mm}
  \subfloat{\includegraphics[width=0.45\linewidth]{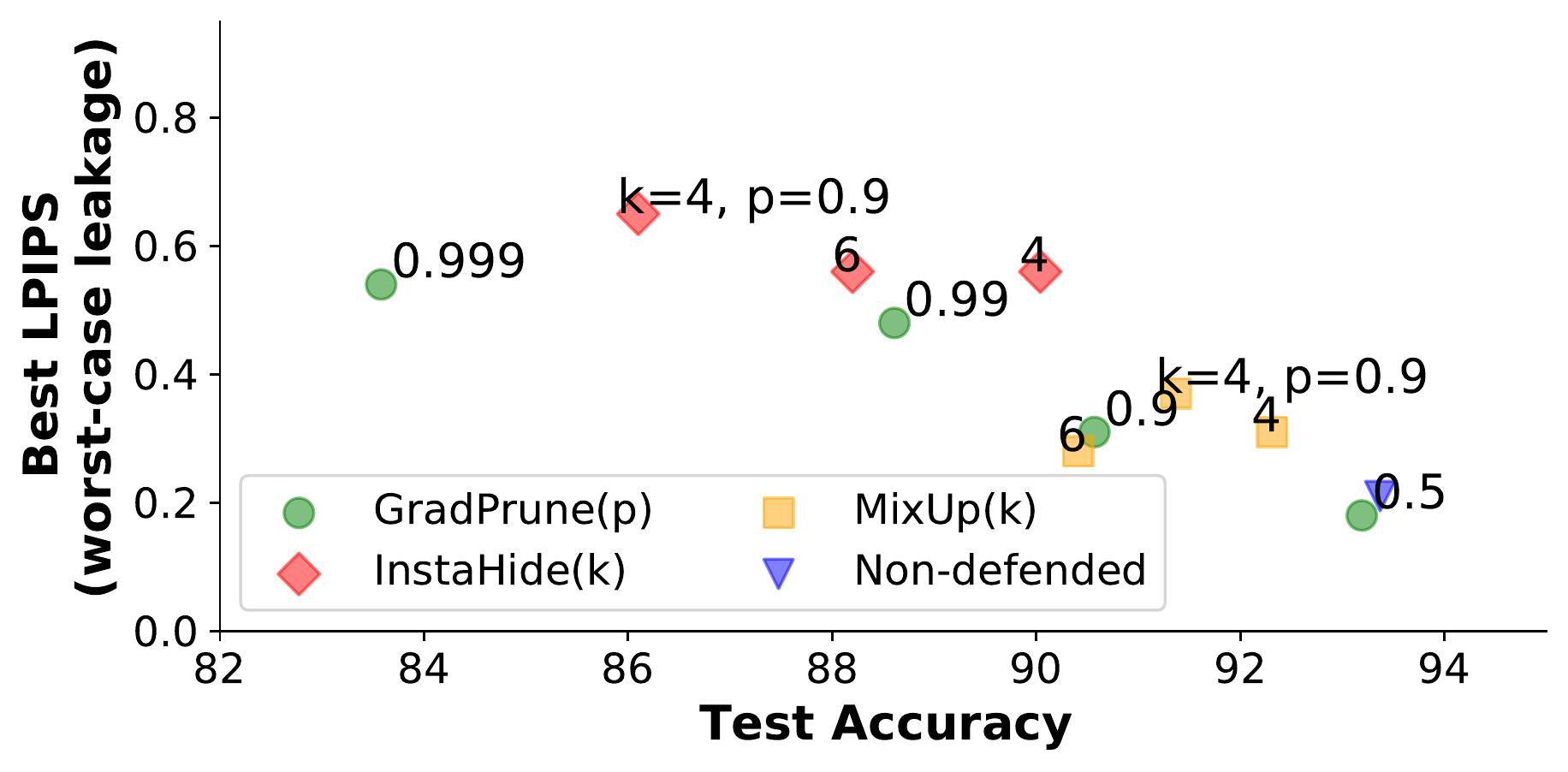}}
  \caption{\small Utility-security trade-off of different defenses. We train the ResNet-18 model on the whole CIFAR-10 dataset, and report the averaged test accuracy and running time of 5 independent runs. We evaluate the attack on a subset of 50 CIFAR-10 images, and report the LPIPS score ($\downarrow$: lower values suggest more privacy leakage). We mark the least-leakage defense measured by the metric in \best{green}.} 
  \label{tab:exp_summary}
  \vspace{-5mm}
\end{table}

\subsection{Performance of defense methods} 
\label{sec:exp_single}

We summarize the performance of each defense in Table~\ref{tab:exp_summary},  and visualize reconstructed images in Figure~\ref{fig:vis_recon}. We report the averaged and the best results for the metric of reconstruction quality, as a proxy for average-case and worst-case privacy leakage.

\paragraph{No defense.} Without any defense, when batch size is 1, the attack can recover images  well from the gradient. Increasing the batch size makes it difficult to recover well, but the recovered images are visually similar to the originals (see Figure~\ref{fig:vis_recon}).

\paragraph{Gradient pruning (GradPrune).}
Figure~\ref{fig:vis_recon} shows that as the pruning ratio $p$ increases, there are more artifacts in the reconstructions. However, the reconstructions are still recognizable even when the pruning ratio $p=0.9$, thus the previous suggestion of using $p=0.7$ by \citep{zhu2020deep} is no longer safe against the state-of-the-art attack. Our results suggest that, for CIFAR-10, defending the strongest attack with gradient pruning may require the pruning ratio $p\geq 0.999$. As a trade-off, such a high pruning ratio would introduce an accuracy loss of around $10\%$ (see Table~\ref{tab:exp_summary}).

\paragraph{MixUp.} MixUp introduces a small computational overhead to training. MixUp with $k=4$ only has a minor impact (\textasciitilde$2\%$) on test accuracy, but it is not sufficient to defend the gradient inversion attack (see Figure~\ref{fig:vis_recon}). Increasing $k$ from $4$ to $6$ slightly reduces the leakage, however, the reconstruction is still highly recognizable.  This suggests that MixUp alone may not be a practical defense against the state-of-the-art gradient inversion attack.

\paragraph{Intra-InstaHide.} Intra-InstaHide with $k=4$ incurs an extra \textasciitilde$2\%$ accuracy loss compared with MixUp, but it achieves better defense performance: when batch size is 32, there are obvious artifacts and color shift in the reconstruction (see Figure~\ref{fig:vis_recon}). However, with batch size 32,  Intra-InstaHide alone also cannot defend the state-of-the-art gradient inversion, as structures of private images are still vaguely identifiable in reconstructions.

Appendix~\ref{sec:app_exp} provides the whole reconstructed dataset under MixUp and Intra-InstaHide.

\subsection{Performance of combined defenses} 
\label{sec:exp_combine}

We notice that two types of defenses (i.e perturbing gradient and encoding inputs) are complementary to each other, which motivates an evaluation of combining gradient pruning with MixUp or  Intra-InstaHide.

As shown in Figure~\ref{fig:vis_recon}, when the batch size is 32, combining  Intra-InstaHide ($k=4$) with gradient pruning ($p=0.9$) makes the reconstruction almost unrecognizable. The combined defense yields a higher LPIPS score than using gradient pruning with $p=0.999$, but introduces a smaller accuracy loss (\textasciitilde$7\%$ compared with a no-defense pipeline). 

Note that our evaluation uses {\em the strongest attack} and {\em relatively small batch sizes}. As shown in Appendix~\ref{sec:app_exp}, invalidating assumptions in Section~\ref{sec:assumption} or increasing the batch size may hinder the attack with an even weaker defense (e.g. with a lower $p$, or smaller $k$), which gives a better accuracy.

\subsection{Time estimate for end-to-end recovery of a single image}
\label{sec:exp_estimate}

Table~\ref{tab:time_estimate} shows time estimates for the end-to-end recovery of a single image in a Federated learning setting with GradPrune or InstaHide defense.  
We do not estimate for MixUp since it has been shown to be a weak defense (see Section~\ref{sec:exp_single}). %We consider the case where the whole dataset is recovered.

Our time estimates consider three fairly small dataset sizes.  The largest size in our estimate is a small fraction of a dataset of ImageNet scale.  We consider a client holds a dataset of $N$ private images and participates in Federated learning, which trains a ResNet-18 model with batch size $b=128$. Assumes that the resolution of the client's data is $32\times 32 \times 3$. If the attacker uses a single NVIDIA GeForce RTX 2080 Ti GPU as his computation resource, and runs gradient inversion with 10,000 iterations of optimization,  then $t$, the running time for attacking a single batch is \textasciitilde$0.25$ GPU hours (batch size $b$ has little impact on the attack's running time, but a larger $b$ makes the attack less effective).

\paragraph{Non-defended and gradient pruning.} Recovering a single image in a non-defended pipeline (or a pipeline that applies gradient pruning alone as the defense) only requires the attacker to invert gradient of a single step of training, which takes time $t$.

\paragraph{InstaHide.} When InstaHide is applied, the current attack~\citep{carlini2020attack} suggests that recovering a single image would involve recovering the whole dataset first. As discussed in Section~\ref{sec:defense}, Carlini et al.'s attack consists of two steps: 1) recover InstaHide images from gradient of $T$ epochs. This would take $ (NT/b) \times t$ GPU hours. 2) Run the decode attack \citep{carlini2020attack} on InstaHide images to recover the private dataset, which involves:

\begin{enumerate}
    \item[2a]  Train a neural network to detect similarity in recovered InstaHide images. Assume that training the network requires at least $n$ recovered InstaHide images, then collecting these images by running gradient inversion would take $(n/b) \times t$ GPU hours. The training takes $10$ GPU hours according to \citep{carlini2020attack}, so training the similarity network would take $(n/b) \times t + 10$ GPU hours in total.
    % \vspace{-2mm}
    \item[2b] Run the combinotorial algorithm to recover original images. Running time of this step has been shown to be at least quadratic in $m$, the number of InstaHide encodings~\citep{chen2020instahide}. This step takes $1/6$ GPU hours with $m=5\times 10 ^{3}$. Therefore for $m=NT$, the running time is at least $1/6 \times (\frac{NT}{5\times 10 ^{3}})^2$ GPU hours.
\end{enumerate}

In total, an attack on InstaHide in this real-world setting would take $ (NT/b) \times t + (n/b) \times t + 10 + 1/6 \times (\frac{NT}{5\times 10 ^{3}})^2$ GPU hours. We use $T=50$ (used by \citep{carlini2020attack}), $n=10,000$ and give estimate in Table~\ref{tab:time_estimate}. As shown, when InstaHide is applied on a small dataset ($N=5,000$), the end-to-end recovery of a single image takes $>3,000\times$ longer than in a no-defense pipeline or GradPrune pipeline; when InstaHide is applied on a larger dataset ($N=500,000$), the computation cost for end-to-end recovery is enormous.

\begin{table}[t]
    \centering
    \footnotesize
    \setlength{\tabcolsep}{10pt}
    \renewcommand{\arraystretch}{0.95}
    \begin{tabular}{|c|c|c|c|}
    \toprule
    {\bf Size of client's dataset ($N$)} & {\bf No defense} & {\bf GradPrune} & {\bf InstaHide} \\
    \midrule
    5,000 & \multirow{3}{*}{0.25} & \multirow{3}{*}{0.25} & 934.48 \\
    50,000 & & & 46,579.01 ($\approx$ 5.5 GPU years)\\
    500,000 & & & 4,215,524.32 ($\approx$ 493.4 GPU years)\\
    \bottomrule
    \end{tabular}
    \vspace{2mm}
    \caption{Time estimates (NVIDIA GeForce RTX 2080 Ti GPU hours) of recovering {\em a single image} from the client's dataset using the state-of-the-art gradient inversion attack~\citep{geiping2020inverting} under different defenses. We assume image resolution of the client's data is $32\times 32 \times 3$. }
    \label{tab:time_estimate}
    \vspace{-5mm}
\end{table}

%% file: conclusion.tex
\section{Conclusions}
\label{sec:discussion}

This paper first points out that some state-of-the-art gradient inversion attacks have made strong assumptions about knowing BatchNorm statistics and private labels.  Relaxing such assumptions can significantly weaken these attacks.

The paper then reports the performance of a set of proposed defenses against gradient inversion attacks, and estimates the computation cost of an end-to-end recovery of a single image in different dataset sizes. Our evaluation shows that InstaHide without mixing with data from a public dataset combined with gradient pruning can defend the state-of-the-art attack, and the estimated time to recover a single image in a medium-size client dataset (e.g. of 500,000 images) is enormous.

Based on our evaluation of the attack by~\citep{geiping2020inverting} and multiple defenses for {\em plain-text} gradients, we have the following observations:

\begin{itemize}
    \item {\em Using BatchNorm layers in your deep net but don't share BatchNorm statistics of the private batch during Federated learning weakens the attack}. We have demonstrated in Section~\ref{sec:assumption} that exposing BatchNorm statistics to the attacker significantly improves the quality of gradient inversion. So a more secure configuration of Federated Learning would be to use BatchNorm layers, but do not share BatchNorm statistics in training, which has been shown feasible in \citep{andreux2020siloed, li2021fedbn}.
    
    \item {\em Using a large batch size weakens the attack; a batch size smaller than 32 is not safe}. We have shown that a larger batch size hinders the attack by making it harder to guess the private labels (Section~\ref{sec:assumption}) and to recover the private images even with correct private labels (Section~\ref{sec:exp}). Our experiments suggest that even with some weak defenses applied, a batch size smaller than 32 is not safe against the strongest gradient inversion attack. 
    
    \item {\em Combining multiple defenses may achieve a better utility-privacy trade-off}. In our experiment, for a batch size of 32, combining InstaHide ($k=4$) with gradient pruning ($p=0.9$) achieves the best utility-privacy trade-off, by making the reconstruction almost unrecognizable at a cost of \textasciitilde$7\%$ accuracy loss (using InstaHide also makes the end-to-end recovery of a single image more computationally expensive). Best parameters would vary for different deep learning tasks, but we strongly encourage Federated learning participants to explore the possibility of combining multiple defensive mechanisms, instead of only using one of them. 
    
\end{itemize}

We hope to extend our work by including  evaluation of defenses for high-resolution images, the attack by \citep{yin2021see}  (when its implementation becomes available), and more defense mechanisms  including those rely on adding noise to gradients.

%% file: app.tex
\appendix

\section{Experimental details and more results}
\label{sec:app_exp}
We run all the experiments on Nvidia RTX 2080 Ti GPUs and V100 GPUs. Table~\ref{tab:app_testbed} summarizes the set of images used in each figure or table in the main paper.  

\captionsetup[table]{font=small}
\begin{table}[H]
    \small
    \centering
    \begin{tabular}{|p{2.5cm}|p{10cm}|}
    \toprule
         {\bf Figure/Table} & {\bf Comments}	\\
    \midrule
        Figure~\ref{fig:BN_var}a & We’ve tuned hyperparams for the attack (see Appendix~\ref{sec:app_hyperparam}) and carried out evaluations on the whole CIFAR-subset. The first sampled batch of size 16 from CIFAR-subset was used in Figure~\ref{fig:BN_var}a to demonstrate the quality of recovery for low-resolution images when BatchNorm statistics are not assumed to be known.  \\
        \midrule
        Figure~\ref{fig:BN_var}b & We’ve tuned hyperparams for the attack (see Appendix~\ref{sec:app_hyperparam}) and carried out evaluations on the whole ImageNet-subset. The best-reconstructed image in ImageNet-subset was used in Figure 1b to demonstrate the quality of recovery for high-resolution images when BatchNorm statistics are not assumed to be known.\\
        \midrule
        Figure~\ref{fig:batch_label_dist} & Percentages of class labels per batch were evaluated over the entire CIFAR10 dataset, for a random seed.	\\
        \midrule
        Figure~\ref{fig:reconstructed_labels} & The first sampled batch of size 16 was used in Figure~\ref{fig:reconstructed_labels} to demonstrate the quality of recovery when labels are not assumed to be known.	\\
        \midrule
        Table~\ref{tab:exp_summary} and Figure~\ref{fig:vis_recon} & We’ve tuned hyperparams for the attack and carried out evaluations on the whole CIFAR-subset. Table~\ref{tab:exp_summary} summarizes the performance of the attack on the whole CIFAR-subset and  Figure~\ref{fig:vis_recon} shows example images.\\
    \bottomrule
    \end{tabular}
    \caption{Summary of experimental testbed for each evaluation.}
    \label{tab:app_testbed}
\end{table}

\subsection{Hyper-parameters}
\label{sec:app_hyperparam}

\paragraph{Training.} For all experiments, we train ResNet-18 for 200 epochs, with a batch size of 128. We use SGD with momentum 0.9 as the optimizer. The initial learning rate is set to 0.1 by default, except for gradient pruning with $p=0.99$ and $p=0.999$. where we set the initial learning rate to 0.02. We decay the learning rate by a factor of 0.1 every 50 epochs.

\begin{figure}[H]
\captionsetup[subfigure]{labelfont=scriptsize, textfont=tiny}
    \centering
    \subfloat[Original]{\includegraphics[width=0.12\textwidth]{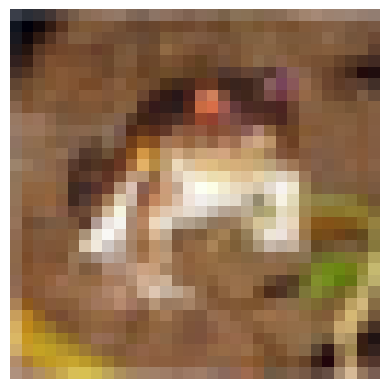}}
    \subfloat[$\alpha_{\rm TV}$=0]{\includegraphics[width=0.12\textwidth]{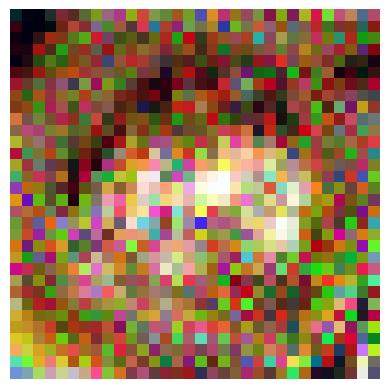}}
    \subfloat[$\alpha_{\rm TV}$=1e-3]{\includegraphics[width=0.12\textwidth]{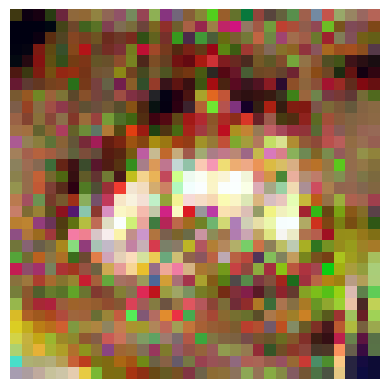}}
    \subfloat[$\alpha_{\rm TV}$=5e-3]{\includegraphics[width=0.12\textwidth]{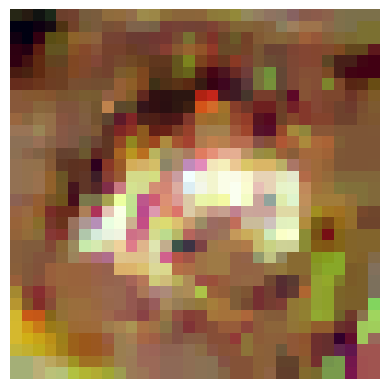}}
    \subfloat[$\alpha_{\rm TV}$=1e-2]{\includegraphics[width=0.12\textwidth]{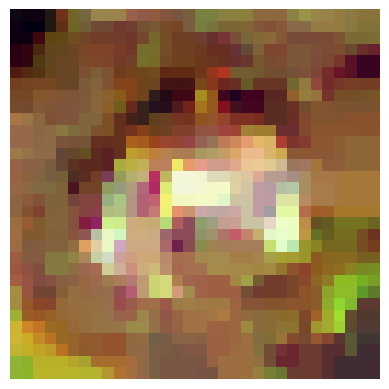}}
    \subfloat[$\alpha_{\rm TV}$=5e-2]{\includegraphics[width=0.12\textwidth]{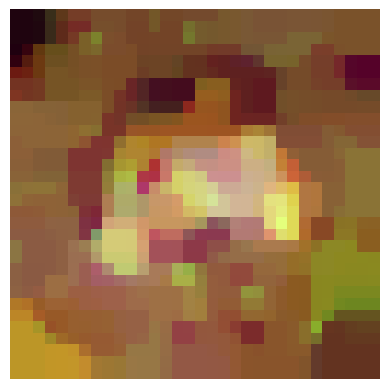}}
    \subfloat[$\alpha_{\rm TV}$=1e-1]{\includegraphics[width=0.12\textwidth]{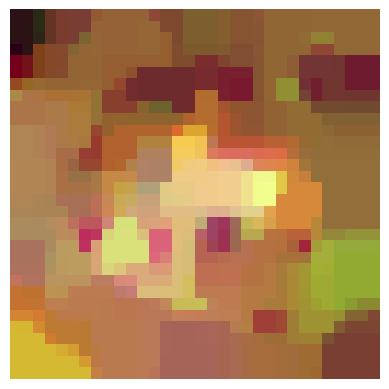}}
    \subfloat[$\alpha_{\rm TV}$=5e-1]{\includegraphics[width=0.12\textwidth]{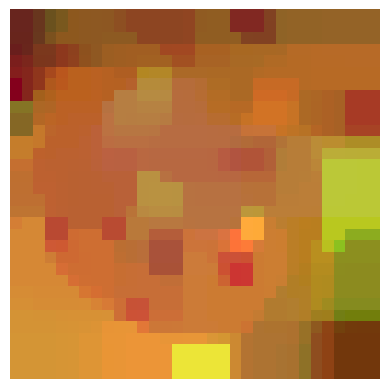}}
    
    \caption{Attacking a single CIFAR-10 images in $\rm BN_{exact}$ setting, with different coefficients for the total variation regularizer ($\alpha_{\rm TV}$'s). $\alpha_{\rm TV}$=1e-2 gives the best reconstruction.}
    \label{fig:BN_tv_tune}
\end{figure}

\begin{figure}[H]
\vspace{-5mm}
\captionsetup[subfigure]{labelfont=scriptsize, textfont=tiny}
    \centering
    \subfloat[Original]{\includegraphics[width=0.16\textwidth]{imgs/assumptions/BN/original.png}}
    \subfloat[$\alpha_{\rm BN}$=0]{\includegraphics[width=0.16\textwidth]{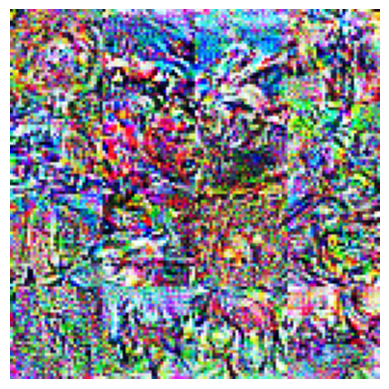}}
    \subfloat[$\alpha_{\rm BN}$=5e-4]{\includegraphics[width=0.16\textwidth]{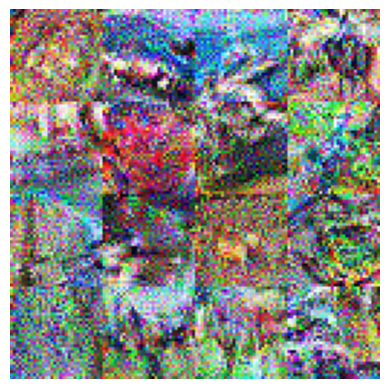}}
    \subfloat[$\alpha_{\rm BN}$=1e-3]{\includegraphics[width=0.16\textwidth]{imgs/assumptions/BN/reconstructed_train_train_bn=1e-3.png}}
    \subfloat[$\alpha_{\rm BN}$=5e-3]{\includegraphics[width=0.16\textwidth]{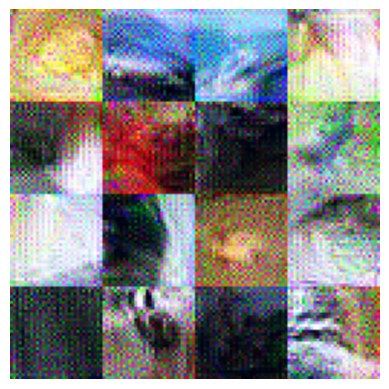}}
    \subfloat[$\alpha_{\rm BN}$=1e-2]{\includegraphics[width=0.16\textwidth ]{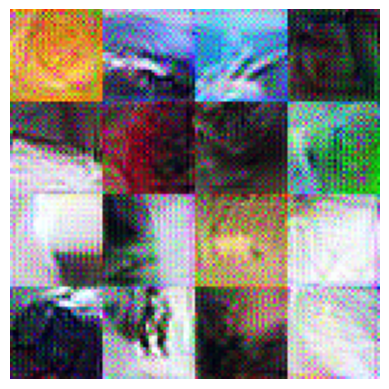}}
    \caption{Attacking a batch of 16 CIFAR-10 images in $\rm BN_{infer}$ setting, with different coefficients for the BatchNorm regularizer ($\alpha_{\rm BN}$'s). $\alpha_{\rm TV}$=1e-3 gives the best reconstruction.}
    \label{fig:BN_reg_tune}
\end{figure}

\paragraph{The attack.}  We report the performance under different $\alpha_{\rm TV}$'s (Figure~\ref{fig:BN_tv_tune}) and $\alpha_{\rm BN}$'s (Figure~\ref{fig:BN_reg_tune}).

\subsection{Details and more results for Section~\ref{sec:assumption}}

\paragraph{Attacking a single ImageNet image.} We launched the attack on ImageNet using the objective function in Eq.~\ref{eq:objective}, where $\alpha_{\rm TV}=0.1$, $\alpha_{\rm BN}=0.001$. We run the attack for 24,000 iterations using Adam optimizer, with initial learning rate 0.1, and decay the learning rate by a factor of $0.1$ at 
$3/8,5/8,7/8$ of training. We rerun the attack 5 times and present the best results measured by LPIPS in Figure~\ref{fig:BN_var}.

\begin{figure}[H]
\captionsetup[subfigure]{font=small}
  \centering
  \subfloat{\includegraphics[width=\textwidth]{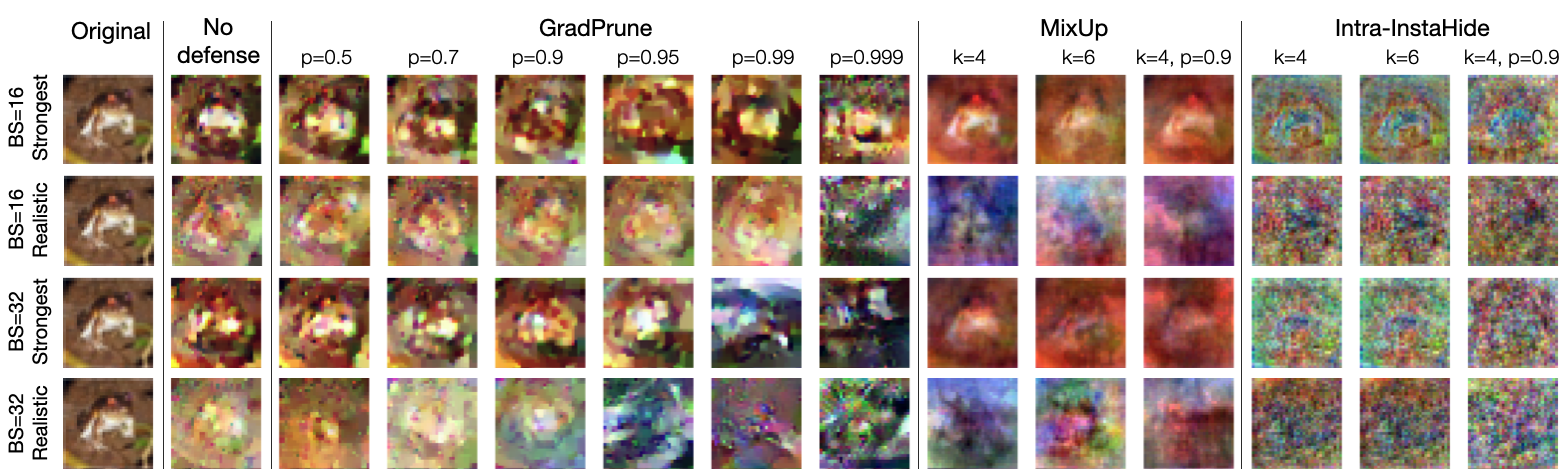}}
  \caption{Reconstruction results under different defenses for a more realistic setting (when the attacker knows private labels but does not know BatchNorm statistics). We also present the results for the strongest attack from Figure~\ref{fig:vis_recon} for comparison. Using Intra-InstaHide with $k=4$ and batch size 32 seems to stop the realistic attack.}
  \label{fig:vis_recon_realistic}
\end{figure}

\paragraph{Qualitative and quantitative results for a more realistic attack.} We also present results of a more realistic attack in Table~\ref{tab:exp_summary_realistic} and Figure~\ref{fig:vis_recon_realistic}, where the attacker does {\em not} know BatchNorm statistics but knows the private labels. We assume the private labels to be known in this evaluation, because for those batches whose distribution of labels is uniform, the restoration of labels should still be quite accurate~\citep{yin2021see}.
As shown, in the evaluated setting, the attack is no longer effective when the batch size is 32 and Intra-InstaHide with $k=4$ is applied. The accuracy loss to stop the realistic attack is only around $3\%$ (compared to around $7\%$ to stop the strongest attack) .

\captionsetup[table]{font=small}
\begin{table}[H] 
  \scriptsize
  \setlength{\tabcolsep}{2.6pt}
  \renewcommand{\arraystretch}{0.95}
  \begin{tabular}{|l|c|c|c|c|c|c|c|c|c|c|c|c|c|c|c|c|}
  \toprule
   &  \multirow{2}{*}{\bf None} & \multicolumn{6}{c|}{\multirow{2}{*}{\bf GradPrune ($p$)}} & \multicolumn{2}{c|}{\multirow{2}{*}{\bf MixUp ($k$)}} & \multicolumn{2}{c|}{\multirow{2}{*}{\bf Intra-InstaHide ($k$)}} & \multicolumn{2}{c|}{\bf GradPrune ($p=0.9$)}\\
   & & \multicolumn{6}{c|}{} & \multicolumn{2}{c|}{} & \multicolumn{2}{c|}{} & {\bf  + MixUp } & {\bf  + Intra-InstaHide}\\
  \midrule
   {\bf Parameter}  & - & 0.5 & 0.7 & 0.9 & 0.95 & 0.99 & 0.999 & 4 & 6 & 4 & 6 & $k=4$ & $k=4$ \\
   \midrule
   {\bf Test Acc.} & 93.37 & 93.19 & 93.01 & 90.57 & 89.92 & 88.61 & 83.58 &  92.31 & 90.41 & 90.04 & 88.20 & 91.37 & 86.10 \\
   \midrule
  {\bf Time (train)} & $1\times$ & \multicolumn{6}{c|}{$1.04\times$} & \multicolumn{2}{c|}{$1.06\times$} & \multicolumn{2}{c|}{$1.06\times$} & \multicolumn{2}{c|}{$1.10\times$} \\
  \midrule
  \multicolumn{14}{|c|}{\bf Attack batch size $= 16$, the strongest attack} \\
  \midrule
  {\bf Avg. LPIPS $\downarrow$}  & 0.41  & 0.41  & 0.42  & 0.46  & 0.48  & 0.50  & 0.55         & 0.50  & 0.49  & 0.69  & 0.69  & 0.62  & \best{0.73}\\
  {\bf Best LPIPS $\downarrow$}  & 0.21  & 0.22  & 0.27  & 0.29  & 0.30  & 0.29  & 0.48         & 0.31  & 0.28  & 0.56  & 0.56  & 0.37  & \best{0.65}\\
  {(LPIPS std.)}                 & 0.09  & 0.08  & 0.07  & 0.06  & 0.06  & 0.06  & 0.04         & 0.10  & 0.10  & 0.06  & 0.07  & 0.10  & 0.05\\
  \midrule
   \multicolumn{14}{|c|}{\bf Attack batch size $= 16$, attacker knows private labels but does not know BatchNorm statistics} \\
   \midrule
   {\bf Avg. LPIPS $\downarrow$}  & 0.49 & 0.51 & 0.48 & 0.51 & 0.52 & 0.56 & 0.60 & 0.71 & 0.71 & \best{0.75} & \best{0.75} & 0.74 &  0.74\\
   {\bf Best LPIPS $\downarrow$}  & 0.30 & 0.33 & 0.31 & 0.33 & 0.34 & 0.39 & 0.44 & 0.48 & 0.53 & \best{0.65} & 0.63 & 0.61 &  0.63\\
   {(LPIPS std.)}                 & 0.08 & 0.09 & 0.08 & 0.08 & 0.07 & 0.07 & 0.05 & 0.08 & 0.07 & 0.04 & 0.05 & 0.08 &  0.05\\
   \midrule
   \multicolumn{14}{|c|}{\bf Attack batch size $= 32$, the strongest attack} \\
  \midrule
  {\bf Avg. LPIPS $\downarrow$}  & 0.45  & 0.46  & 0.48  & 0.52  & 0.54  & 0.58  & 0.63         & 0.50  & 0.49  & 0.69  & 0.69  & 0.62  & \best{0.73}\\
   {\bf Best LPIPS $\downarrow$}  & 0.18  & 0.18  & 0.22  & 0.31  & 0.43  & 0.48  & 0.54         & 0.31  & 0.28  & 0.56  & 0.56  & 0.37  & \best{0.65}\\
   {(LPIPS std.)}                 & 0.11  & 0.11  & 0.09  & 0.07  & 0.05  & 0.04  & 0.04         & 0.10  & 0.10  & 0.06  & 0.07  & 0.10  & 0.05\\
    \midrule
   \multicolumn{14}{|c|}{\bf Attack batch size $= 32$, attacker knows private labels but does not know BatchNorm statistics} \\
   \midrule
   {\bf Avg. LPIPS $\downarrow$}  & 0.48 & 0.50 & 0.53 & 0.53 & 0.55 & 0.60 & 0.63 & 0.73 & 0.72 & 0.76 & 0.76 & 0.76 & \best{0.77} \\
   {\bf Best LPIPS $\downarrow$}  & 0.29 & 0.32 & 0.32 & 0.31 & 0.40 & 0.41 & 0.55 & 0.63 & 0.60 & \best{0.68} & 0.63 & 0.66 & 0.65\\
   {(LPIPS std.)}                 & 0.08 & 0.07 & 0.07 & 0.08 & 0.08 & 0.06 & 0.04 & 0.06 & 0.06 & 0.04 & 0.05 & 0.06 & 0.05\\
  \bottomrule
  \end{tabular}
  \vspace{2mm}
%   \subfloat{\includegraphics[width=0.98\textwidth]{imgs/Compare_16_32.png}}
  \caption{\small Utility-security trade-off of different defenses for a more realistic setting (when the attacker knows private labels but does not know BatchNorm statistics). We also present the results for the strongest attack from Table~\ref{tab:exp_summary} for comparison. We evaluate the attack on 50 CIFAR-10 images and report the LPIPS score ($\downarrow$: lower values suggest more privacy leakage).
  We mark the least-leakage defense measured by the metric in \best{green}.} 
  \label{tab:exp_summary_realistic}
\end{table}

\subsection{More results for the strongest attack}

\begin{figure}[H]
\captionsetup[subfigure]{font=small}
  \centering
  \vspace{-12mm}
  \subfloat[Batch size $=1$]{\includegraphics[width=\linewidth]{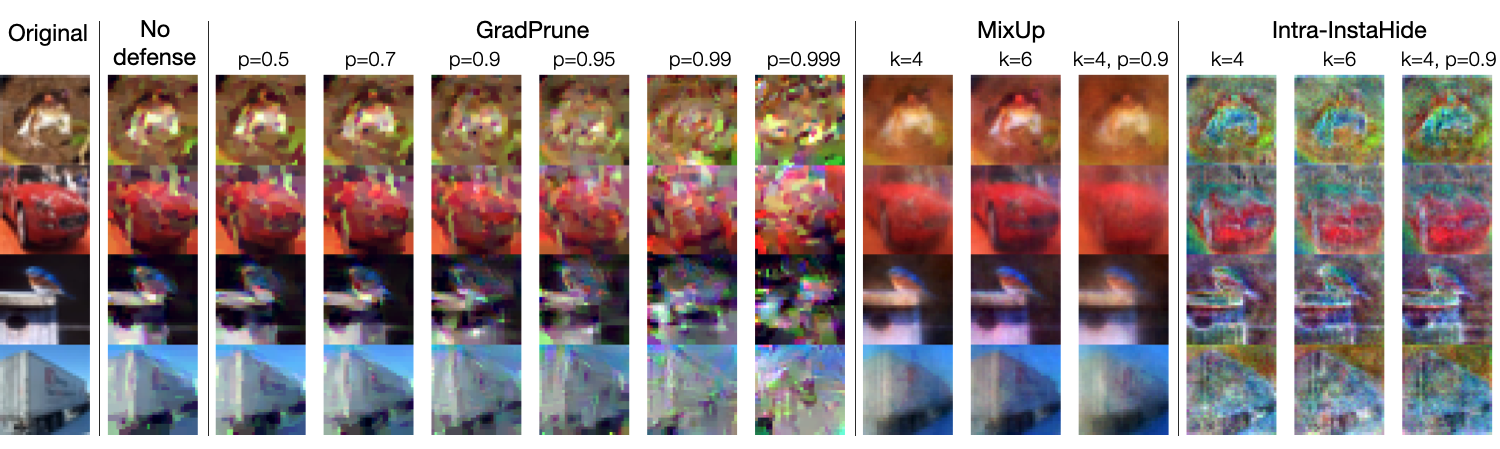}}\\
  \vspace{-3mm}
  \subfloat[Batch size $=16$]{\includegraphics[width=\linewidth]{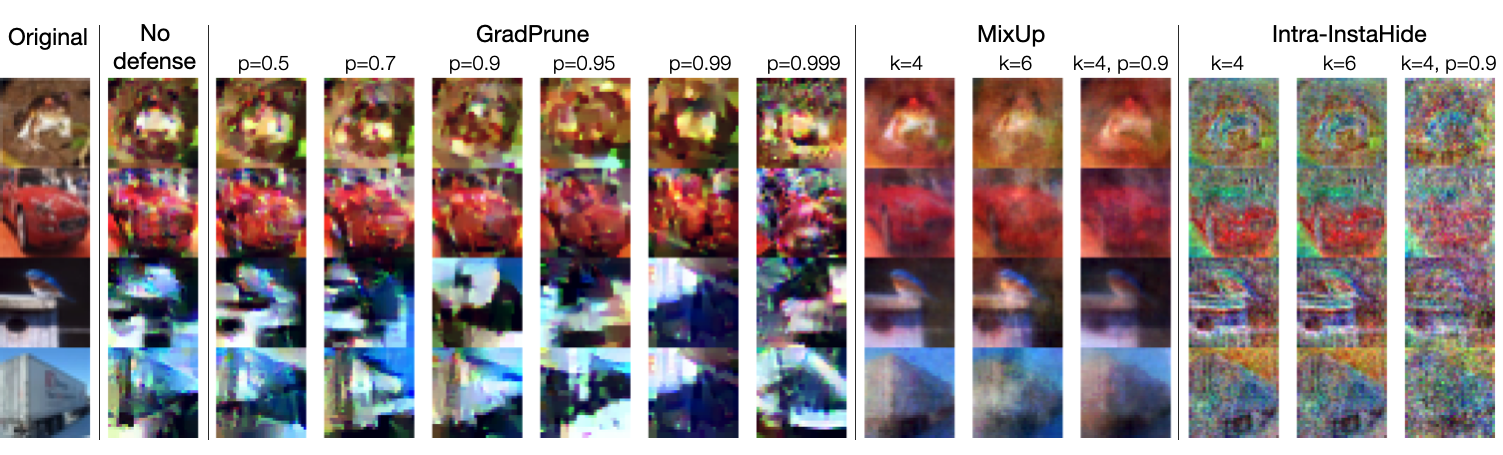}}\\
  \vspace{-3mm}
  \subfloat[Batch size $=32$]{\includegraphics[width=\linewidth]{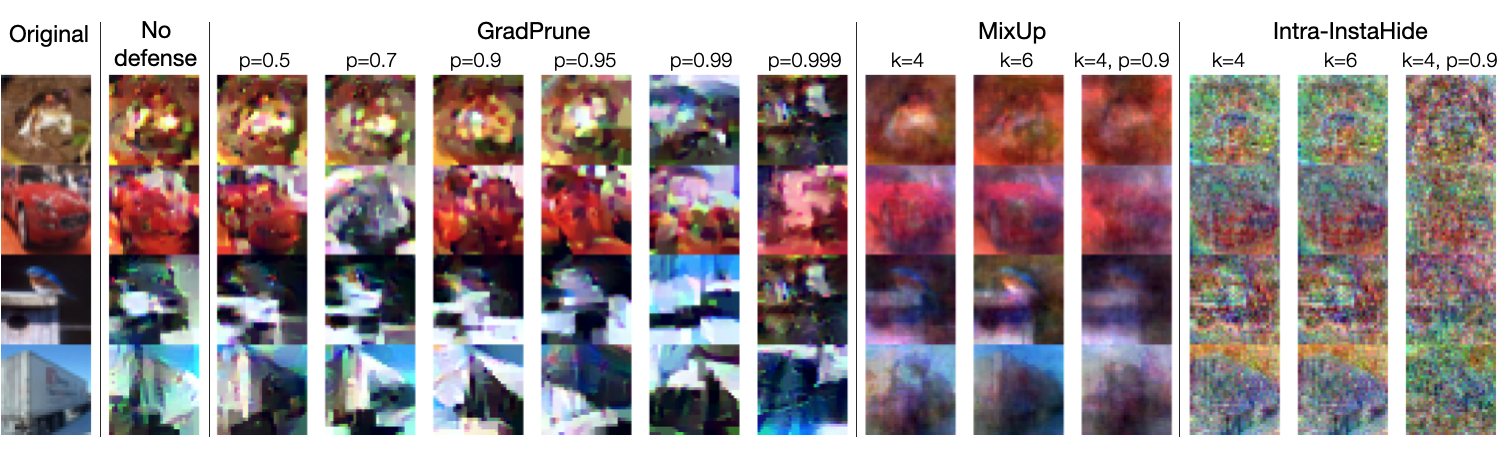}}\\
  \vspace{-2mm}
  \caption{Reconstruction results under different defenses with batch size 1 (a), 16 (b) and 32 (c). Full version of Figure~\ref{fig:vis_recon}.}
  \label{fig:vis_recon_full}
  \vspace{-2mm}
\end{figure}

\paragraph{Full version of Figure~\ref{fig:vis_recon}.} Figure~\ref{fig:vis_recon_full} provides more examples for reconstructed images by the strongest attack under different defenses and batch sizes.

\paragraph{Results with MNIST dataset.} We’ve repeated our main evaluation of defenses and attacks (Table~\ref{tab:exp_summary}) on MNIST dataset~\citep{deng2012mnist} with a simple 6-layer ConvNet model. Note that the simple ConvNet does not contain BatchNorm layers. We evaluate the following defenses on the MNIST dataset with a 6-layer ConvNet architecture against the strongest attack (private labels known):

\begin{itemize}
    \item GradPrune (gradient pruning): gradient pruning sets gradients of small magnitudes to zero. We vary the pruning ratio $p$ in \{0.5, 0.7, 0.9, 0.95, 0.99, 0.999, 0.9999\}.
    \item MixUp: we vary $k$ in \{4,6\}, and set the upper bound of a single coefficient to 0.65 (coefficients sum to 1).
    \item Intra-InstaHide: we vary $k$ in \{4,6\}, and set the upper bound of a single coefficient to 0.65 (coefficients sum to 1). 
    \item A combination of GradPrune and MixUp/Intra-InstaHide.
\end{itemize}

We run the evaluation against the strongest attack and batch size 1 to estimate the upper bound of privacy leakage. Specifically, we assume the attacker knows private labels, as well as the indices of mixed images and mixing coefficients for MixUp and Intra-InstaHide. 

\begin{figure}[t]
    \centering
    \includegraphics[width=0.95\linewidth]{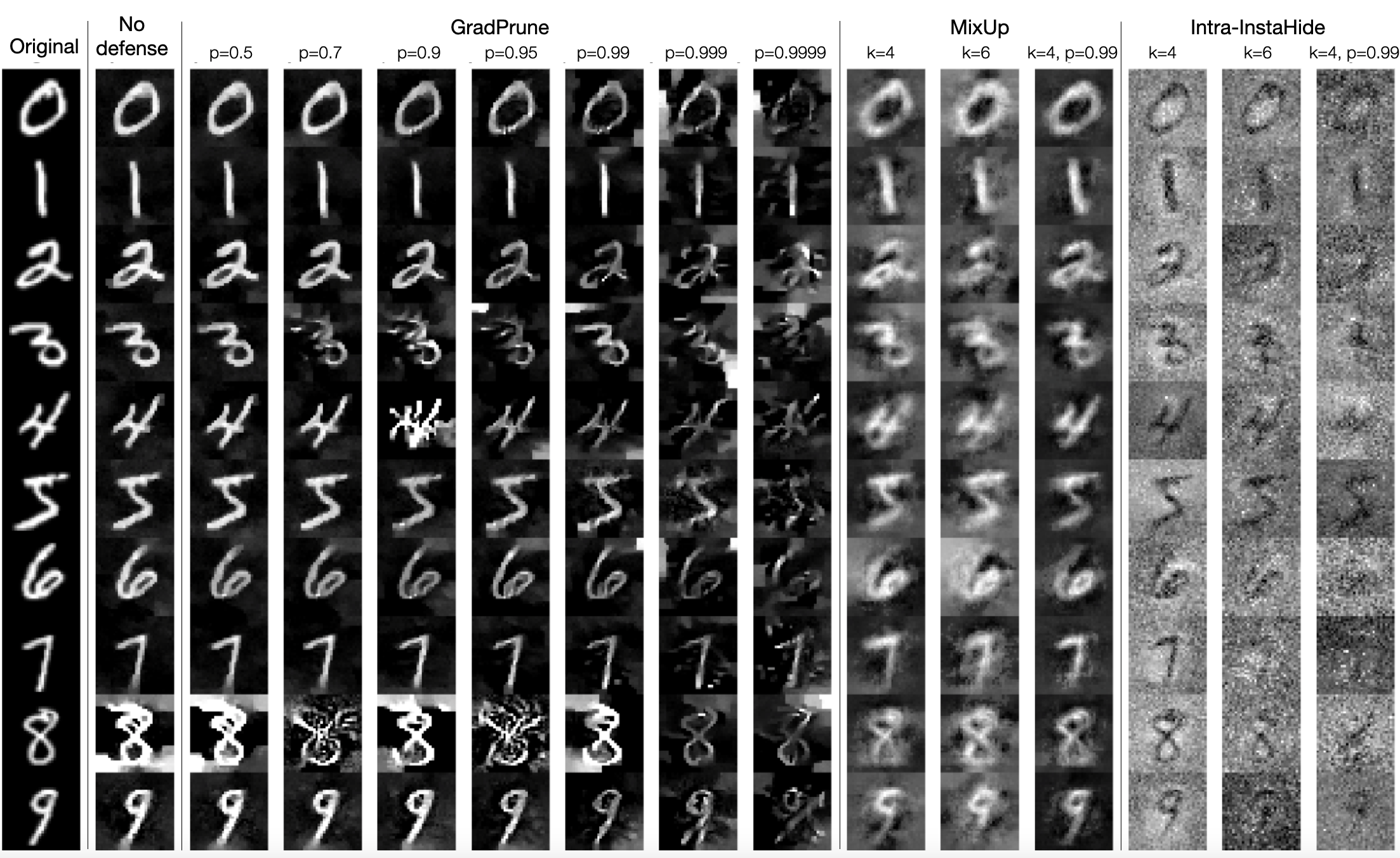}
    \caption{Reconstruction results of MNIST digits under different defenses with the strongest atttack and batch size 1.}
    \label{fig:vis_recon_MNIST}
    \vspace{-5mm}
\end{figure}

For MNIST with a simple 6-layer ConvNet, defending the strongest attack with gradient pruning may require the pruning ratio $p\geq 0.9999$. MixUp with $k=4$ or $k=6$ are not sufficient to defend the gradient inversion attack. Combining MixUp ($k=4$) with gradient pruning ($p=0.99$) improves the defense, however, the reconstructed digits are still highly recognizable. Intra-InstaHide alone ($k=4$ or $k=6$) gives a bit better defending performance than MixUp and GradPrune. Combining InstaHide ($k=4$) with gradient pruning ($p=0.99$) further improves the defense and makes the reconstruction almost unrecognizable.

\subsection{More results for encoding-based defenses}
\begin{figure}[H]
    \centering
    \includegraphics[width=0.95\textwidth]{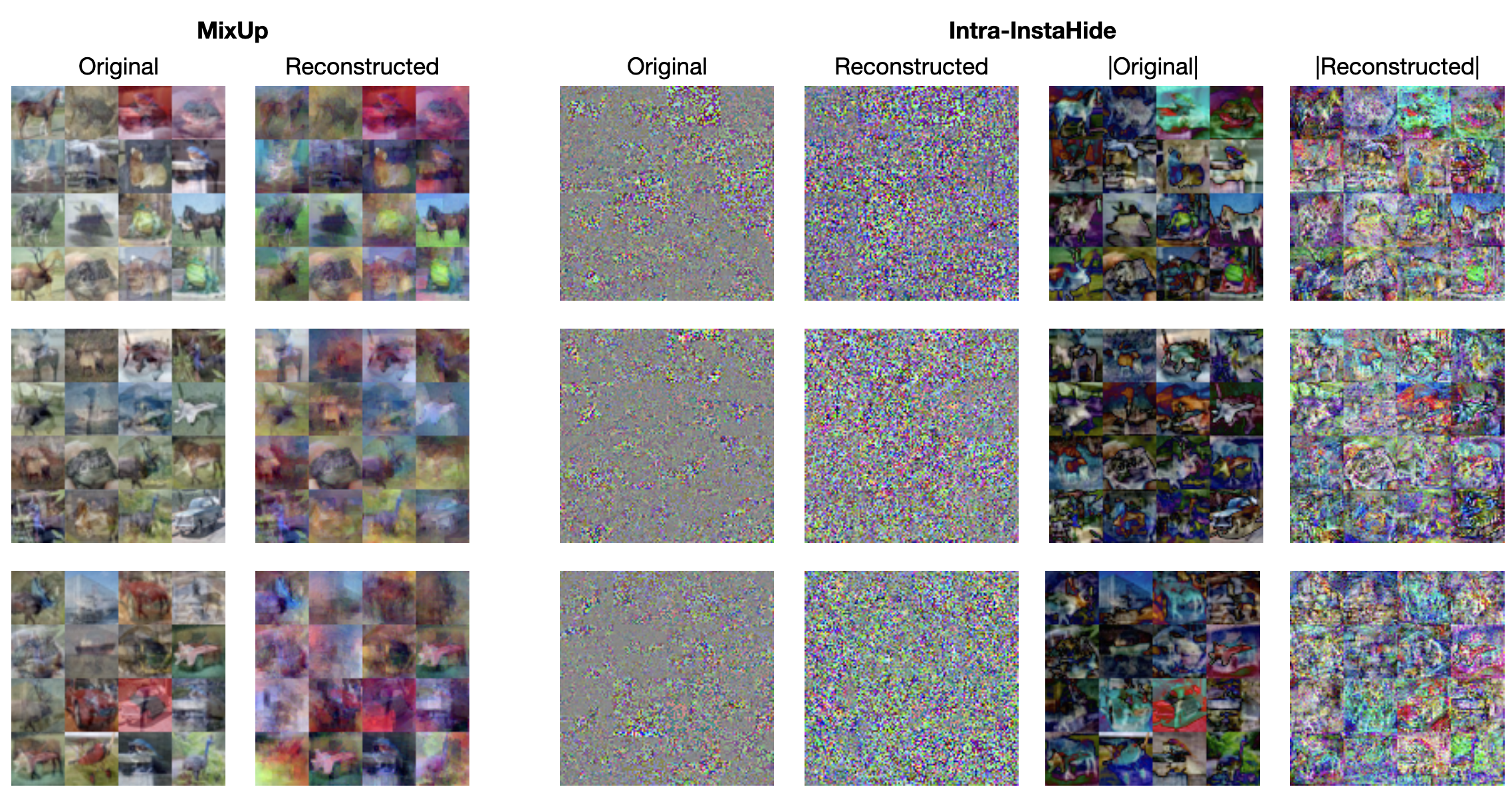}
    \caption{Original and reconstructed batches of 16 images under MixUp and Intra-InstaHide defenses. We visualize both the original and the absolute images for the Intra-InstaHide defense. Intra-InstaHide makes pixel-wise matching harder.}
    \label{fig:mixup_vs_instahide}
    \vspace{-5mm}
\end{figure}

\begin{figure}[H]
\captionsetup[subfigure]{labelfont=scriptsize, textfont=tiny}
    \centering
    \subfloat[Original]{\includegraphics[width=0.23\textwidth]{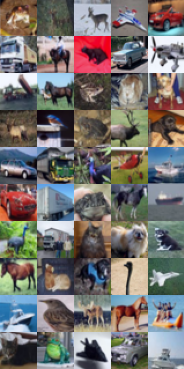}} \hspace{1mm}
    \subfloat[MixUp, $k$=4]{\includegraphics[width=0.23\textwidth]{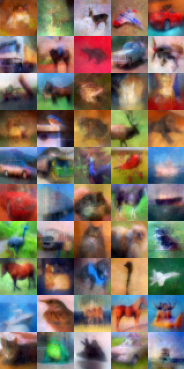}} \hspace{1mm}
    \subfloat[MixUp, $k$=6]{\includegraphics[width=0.23\textwidth]{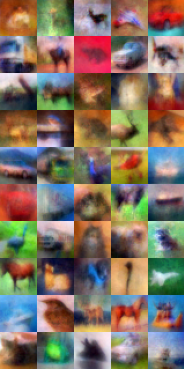}} \hspace{1mm}
    \subfloat[MixUp+GradPrune, $k$=4, $p$=0.9]{\includegraphics[width=0.23\textwidth]{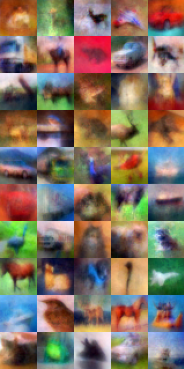}}
    
    \subfloat[Original]{\includegraphics[width=0.23\textwidth]{imgs/decode_res/InstaHide/bs1_k4/originals.png}} \hspace{1mm}
    \subfloat[InstaHide, $k$=4]{\includegraphics[width=0.23\textwidth]{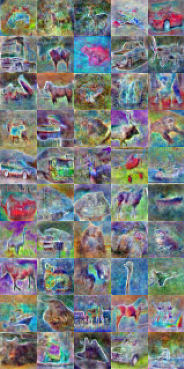}} \hspace{1mm}
    \subfloat[InstaHide, $k$=6]{\includegraphics[width=0.23\textwidth]{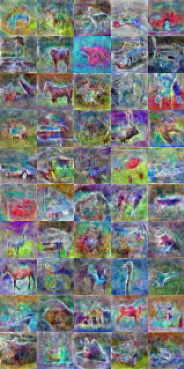}} \hspace{1mm}
    \subfloat[InstaHide+GradPrune, $k$=4, $p$=0.9]{\includegraphics[width=0.23\textwidth]{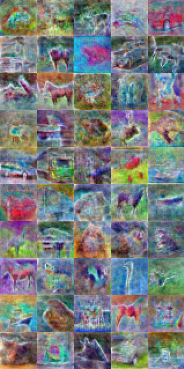}}
    \caption{Reconstrcuted dataset under MixUp and Intra-InstaHide against the strongest attack (batch size is 1).}
    \label{fig:encode_bs1}
    \vspace{-10mm}
\end{figure}

\begin{figure}[H]
\captionsetup[subfigure]{labelfont=scriptsize, textfont=tiny}
    \centering
    \subfloat[Original]{\includegraphics[width=0.23\textwidth]{imgs/decode_res/InstaHide/bs1_k4/originals.png}} \hspace{1mm}
    \subfloat[MixUp, $k$=4]{\includegraphics[width=0.23\textwidth]{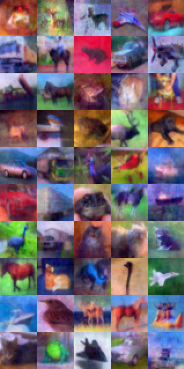}} \hspace{1mm}
    \subfloat[MixUp, $k$=6]{\includegraphics[width=0.23\textwidth]{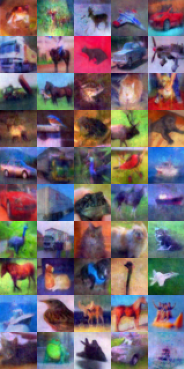}} \hspace{1mm}
    \subfloat[MixUp+GradPrune, $k$=4, p=0.9]{\includegraphics[width=0.23\textwidth]{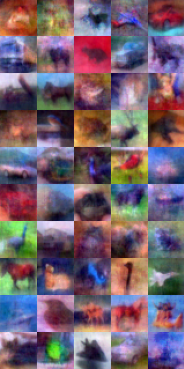}}

    \subfloat[Original]{\includegraphics[width=0.23\textwidth]{imgs/decode_res/InstaHide/bs1_k4/originals.png}} \hspace{1mm}
    \subfloat[InstaHide, $k$=4]{\includegraphics[width=0.23\textwidth]{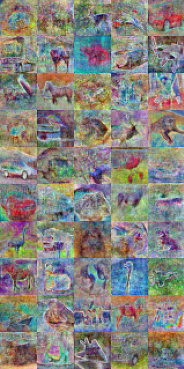}} \hspace{1mm}
    \subfloat[InstaHide, $k$=6]{\includegraphics[width=0.23\textwidth]{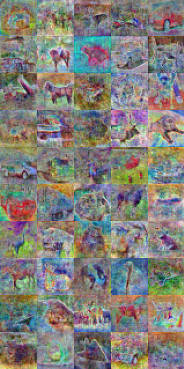}} \hspace{1mm}
    \subfloat[InstaHide+GradPrune, $k$=4, $p$=0.9]{\includegraphics[width=0.23\textwidth]{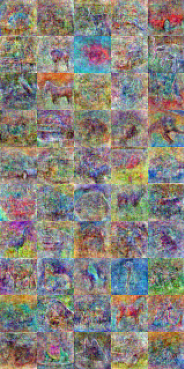}}
    \caption{Reconstrcuted dataset under MixUp and Intra-InstaHide against the strongest attack (batch size is 16).}
    \label{fig:encode_bs16}
\end{figure}

\begin{figure}[H]
\captionsetup[subfigure]{labelfont=scriptsize, textfont=tiny}
    \centering
    \subfloat[Original]{\includegraphics[width=0.23\textwidth]{imgs/decode_res/InstaHide/bs1_k4/originals.png}} \hspace{1mm}
    \subfloat[MixUp, $k$=4]{\includegraphics[width=0.23\textwidth]{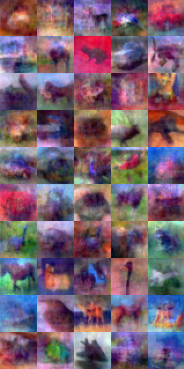}} \hspace{1mm}
    \subfloat[MixUp, $k$=6]{\includegraphics[width=0.23\textwidth]{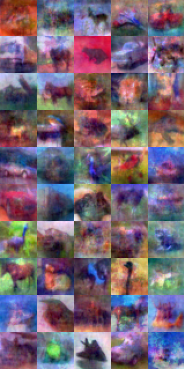}} \hspace{1mm}
    \subfloat[MixUp+GradPrune, $k$=4, $p$=0.9]{\includegraphics[width=0.23\textwidth]{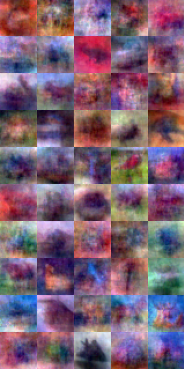}}

    \subfloat[Original]{\includegraphics[width=0.23\textwidth]{imgs/decode_res/InstaHide/bs1_k4/originals.png}} \hspace{1mm}
    \subfloat[InstaHide, $k$=4]{\includegraphics[width=0.23\textwidth]{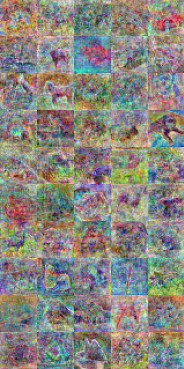}} \hspace{1mm}
    \subfloat[InstaHide, $k$=6]{\includegraphics[width=0.23\textwidth]{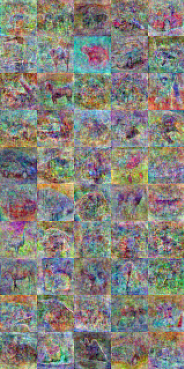}} \hspace{1mm}
    \subfloat[InstaHide+GradPrune, $k$=4, $p$=0.9]{\includegraphics[width=0.23\textwidth]{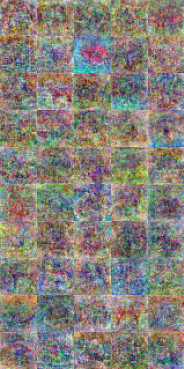}}
    \caption{Reconstrcuted dataset under MixUp and Intra-InstaHide against the strongest attack (batch size is 32).}
    \label{fig:encode_bs32}
\end{figure}

We visualize the whole reconstructed dataset under MixUp and Intra-InstaHide defenses with different batch sizes in Figure~\ref{fig:encode_bs1}, \ref{fig:encode_bs16} and \ref{fig:encode_bs32}.  Sample results of the original and the reconstructed batches are provided in Figure~\ref{fig:mixup_vs_instahide}.

\input{problem}

%% file: problem.tex
\newpage

\section{Theoretical Insights for Defenses' Working Mechanism}
\label{sec:app_theory}
In this section, we provide some theoretical insights for the mechanism of each defense.

\subsection{Gradient pruning}
Gradient pruning is a non-oblivious case of applying sketching techniques~\citep{cls19} to compress the gradient vector. Usually, if we only observe the vector after sketching, it is hard to recover the original vector unless certain assumptions of the vector itself and the sketching technique have been made. Therefore, gradient pruning prevents the attacker from seeing the original gradient and make the inversion harder. 

\subsection{MixUp and InstaHide} Intuitively, MixUp and InstaHide's working mechanism may come from mixing $k$ images in a single encoded image, which appears to be similar to multiplying the batch size by a factor of $k$, thus makes the attack less effective. In Section~\ref{sec:gradient}, we provide theoretical analysis for this intuition, by showing that mixing $k$ images and using a batch size of $k$ are essentially similar, with any neural network that has a fully connected layer as its first layer. 

Another layer of security of InstaHide seems to come from applying random sign-flipping on the mixed images. As mentioned in Section~\ref{sec:exp_setup}, for an InstaHide-encoded image $x \in \R^d$, we apply the total variation regularizer on $|x|$ instead of $x$, which pushes the gap between absolute value of adjacent pixels (i.e., $| |x_j| - |x_{j+1} | |$) to be small. However having $| |x_j| - |x_{j+1}|  | = \delta$ for some small $\delta < 10^{-4}$ does not imply that $|x_j - x_{j+1}| = \delta$; in fact, $|x_j - x_{j+1}|$ can be as large as $1-\delta$.  Therefore, the random sign flipping operation in InstaHide could potentially make the total variation image prior less effective in some sense (see Figure~\ref{fig:mixup_vs_instahide}). 

\iffalse
\paragraph{Mixing up images is (nearly) increasing batch size}
using one image which is generating from mix-up 4 images is somehow equivalent to using batch size that has 4 images.

Say images are $x_1, \cdots, x_4$

Batched gradient looks like this
\begin{align*}
    \sum_{i=1}^4 {\bf 1}_{\langle w, x_i \rangle } x_i
\end{align*}

MixUp gradient looks 
\begin{align*}
    \sum_{i=1}^4 {\bf 1}_{\langle w, \overline{x} \rangle } x_i
\end{align*}
where $\overline{x} = \sum_{i=1}^4 x_i $
\fi

\subsection{Property of gradient in a small batch}\label{sec:gradient}

The goal of this section is to present the following results,
\begin{lemma}\label{lem:main_gradient}
Given a neural network with ReLU activation function, each row of the gradient of first layer weights is a linear combination of images, i.e.
\begin{align*}
    ( \frac{\partial {\cal L}(W)} { \partial W_1 } )_i = \sum_{j=1}^b \alpha_{i,j} x_i^\top 
\end{align*}
where the $b$ is the number of images in a small batch, $\{x_1, \cdots, x_b \} \in \R^d$ are images in that small batch. 
\end{lemma}
In Section~\ref{sec:gradient_1} and \ref{sec:gradient_2}, we show the above observation holds for one/two-hidden layer neural network. In Section~\ref{sec:gradient_l}, we generalize it to multiple layer neural network.

The standard batched $k$-vector sum can be defined as follows:
\begin{definition}
Give a database $X$ list of vectors $x_1, \cdots, x_n$. Given a list of observations $y_1, \cdots, y_m$ where for each $j \in [m]$, there is a set $S_j$ such that $y_j = \sum_{i \in S_j} x_i$ 
and $|S_j|=b$. We can observe $y_1, \cdots y_m$ but has no access to database, the goal is to recover $S_j$ and the vectors $x_i$ being use, for each $j$.
\end{definition}
The above definition is a mathematical abstraction of MixUp recovery/attack. It can be further generalized to InstaHide, if we only observe the $|y_j|$. We also want to remark that in the above definition, we simplify the formulation by using coefficients $1$ for all vectors. It can be easily generalized to 
settings where random coefficients are assigned to vectors in the database for MixUp/InstaHide. 

Using Lemma~\ref{lem:main_gradient}, we notice that
\begin{lemma}
Under the condition of Lemma~\ref{lem:main_gradient}, given a list of observation of gradients, and the problem recovering images is also a batched vector sum problem. 

\end{lemma}
Thus, gradient attack is essentially an variation of MixUp/Instahide attack.

\subsection{One Hidden Layer}\label{sec:gradient_1}
We consider a one-hidden layer ReLU activated neural network with $m$ neurons in the hidden layer:
\begin{align*}
    f(x) = a^\top \phi(W x )
\end{align*}
where $a\in \R^m$ and $W \in \R^{m \times d}$.
We define objective function $L$ as follows:
\begin{align*}
    L(W) = \frac{1}{2} \sum_{i=1}^n (y_i - f(W,x_i,a))^2
\end{align*}
We can compute the gradient of $L$ in terms of $w_r$
\begin{align*}
    \frac{ \partial L(W) }{ \partial w_r } = \sum_{i=1}^n ( f(W,x_i,a) -y_i ) a_r x_i {\bf 1}_{ \langle w_r, x_i \rangle }
\end{align*}
Let $\wt{x} = \frac{1}{n} \sum_{i=1}^n x_i$,
\begin{align*}
    \frac{ \partial L(W) }{ \partial w_r } = ( f(W, \wt{x}, a) - y ) a_r \wt{x} {\bf 1}_{\langle w_r, \wt{x} \rangle}
\end{align*}
Another version
\begin{align*}
    \frac{ \partial L(W) }{ \partial w_r } = \sum_{i=1}^n ( f(W, x_i, a) - y_i ) \cdot \Big( a_r \wt{x} {\bf 1}_{\langle w_r, \wt{x} \rangle} \Big)
\end{align*}

\subsection{Two Hidden Layers}\label{sec:gradient_2}

Suppose $a \in \R^m$, $V \in \R^{m \times d}, W \in \R^{m \times m}$. The neural network is defined as $f : \R^d \rightarrow \R$, here we slightly deviate from the standard setting and assume the input dimension is $m$, in order to capture the general setting.
\begin{align*}
f(x) = a^{\top} \phi( W \phi ( V x ) )
\end{align*}
Consider the mean square loss
\begin{align*}
L(W, V, a) = \frac{1}{2} \sum_{i=1}^n | f(x_i) - y_i |^2
\end{align*}
The gradient with respect to $W$ is
\begin{align*}
\frac{ \partial L(W, V, a)}{ \partial W } = \sum_{i=1}^{n}(f(x_i) - y_i)  \underbrace{ \diag\{\phi'(W \phi(Vx_i))\} }_{ m \times m }\underbrace{ a }_{m \times 1}  \underbrace{ \phi(Vx_i)^{\top} }_{1 \times m}  
\end{align*}
and the gradient with respect to $V$ is
\begin{align*}
\frac{ \partial L(W, V, a)}{ \partial V } = \sum_{i=1}^{n}(f(x_i) - y_i)\underbrace{\diag\{ \phi'(Vx_i)\}}_{m \times m} \underbrace{W^{\top}}_{m \times m}  \underbrace{\diag\{\phi'(W \phi(Vx_i))\}}_{m \times m} \underbrace{a}_{m \times 1}    \underbrace{x_i^{\top}}_{1 \times d} 
\end{align*}

\subsection{The multi-layers case}\label{sec:gradient_l}

The following multiple layer neural network definition is standard in literature. %should be found in \cite{als19_dnn}.

Consider a $L$ layer neural network with one vector $a\in \R^{m_L}$ and $L$ matrices $W_L \in \R^{m_L \times m_{L-1}}$, $\cdots$, $W_2 \in \R^{m_2 \times m_1}$ and $W_1 \in \R^{m_1 \times m_0}$. Let $m_0 = d$.
In order to write gradient in an elegant way, we define some artificial variables as follows
\begin{align}\label{eq:def_g_h}
g_{i,1} = & ~ W_1 x_i, && h_{i,1} = \phi(W_1 x_i), && \in \R^{m_1} && \forall i \in [n]\notag\\
g_{i,\ell} = & ~ W_{\ell} h_{i,\ell-1}, && h_{i,\ell} = \phi( W_{\ell} h_{i,\ell-1} ), && \in \R^{m_{\ell}} && \forall i \in [n], \forall \ell \in \{2,3,\cdots,L\} \\
%f(W,x_i) = & ~ a^\top h_{i,L}, && && \in \R && \forall i \in [n]\notag \\
D_{i, 1} = & ~ \text{diag} \big(\phi'(W_1 x_i )\big), && && \in \R^{m_1 \times m_1} && \forall i \in [n] \notag\\
D_{i, \ell} = & ~ \text{diag} \big(\phi'(W_{\ell} h_{i, \ell-1})\big), && && \in \R^{m_{\ell} \times m_{\ell}} && \forall i \in [n], \forall \ell \in \{2,3,\cdots,L\} \notag
%\label{eq:nn}
\end{align}
where $\phi(\cdot)$ is the activation function and $\phi'(\cdot)$ is the derivative of activation function.

Let $f : \R^{m_0} \rightarrow \R$ denote neural network function: 
\begin{align*}
    f(W,x) = a^\top \phi( W_L ( \phi ( \cdots \phi(W_1 x ) ) ) )
\end{align*}
Thus using definition of $f$ and $h$, we have
\begin{align*}
    f(W,x_i) =  a^\top h_{i,L}, ~~~ \in \R, ~~~ \forall i \in [n]
\end{align*}

Given $n$ input data points $(x_1 , y_1), (x_2 , y_2) , \cdots (x_n, y_n) \in \R^{d} \times \R$. We define the objective function $\mathcal{L}$ as follows
\begin{align*}
\mathcal{L} (W) = \frac{1}{2} \sum_{i=1}^n ( y_i - f (W,x_i) )^2 .
\end{align*}

We can compute the gradient of ${\cal L}$ in terms of $W_{\ell} \in \R^{m_{\ell} \times m_{\ell-1}}$, for all $\ell \geq 2$
\begin{align}\label{eq:gradient}
\frac{\partial\mathcal{L}(W)}{\partial W_{\ell}} = \sum_{i = 1}^{n} ( f(W, x_i) - y_i) \underbrace{D_{i, \ell}}_{ m_{\ell} \times m_{\ell} } \left( \prod_{k = \ell+1}^{L} \underbrace{W_{k}^{\top}}_{m_{k-1} \times m_k}\underbrace{D_{i, k}}_{ m_{k} \times m_k } \right)  \underbrace{a}_{ m_L \times 1} \underbrace{h_{i, \ell -1}^{\top}}_{1 \times m_{\ell-1}}
\end{align}
Note that the gradient for $W_1 \in \R^{m_1 \times m_0}$ (recall that $m_0 = d$) is slightly different and can not be written by general form. Here is the form
\begin{align}
    \frac{\partial\mathcal{L}(W)}{\partial W_{1}} = \sum_{i = 1}^{n} ( f(W, x_i) - y_i) \underbrace{D_{i, 1}}_{ m_1 \times m_1 } \left( \prod_{k = 2}^{L} \underbrace{W_{k}^{\top}}_{m_{k-1} \times m_k}\underbrace{D_{i, k}}_{ m_{k} \times m_k } \right)  \underbrace{a}_{ m_L \times 1} \underbrace{x_{i}^{\top}}_{1 \times m_{0}}
\end{align}

%% file: main.bbl
\begin{thebibliography}{37}
\providecommand{\natexlab}[1]{#1}
\providecommand{\url}[1]{\texttt{#1}}
\expandafter\ifx\csname urlstyle\endcsname\relax
  \providecommand{\doi}[1]{doi: #1}\else
  \providecommand{\doi}{doi: \begingroup \urlstyle{rm}\Url}\fi

\bibitem[Abadi et~al.(2016)Abadi, Chu, Goodfellow, McMahan, Mironov, Talwar,
  and Zhang]{abadi2016deep}
Martin Abadi, Andy Chu, Ian Goodfellow, H~Brendan McMahan, Ilya Mironov, Kunal
  Talwar, and Li~Zhang.
\newblock Deep learning with differential privacy.
\newblock In \emph{Proceedings of the 2016 ACM SIGSAC conference on computer
  and communications security}, pages 308--318, 2016.

\bibitem[Act(1996)]{hippa}
Accountability Act.
\newblock Health insurance portability and accountability act of 1996.
\newblock \emph{Public law}, 104:\penalty0 191, 1996.

\bibitem[Andreux et~al.(2020)Andreux, du~Terrail, Beguier, and
  Tramel]{andreux2020siloed}
Mathieu Andreux, Jean~Ogier du~Terrail, Constance Beguier, and Eric~W Tramel.
\newblock Siloed federated learning for multi-centric histopathology datasets.
\newblock In \emph{Domain Adaptation and Representation Transfer, and
  Distributed and Collaborative Learning}, pages 129--139. Springer, 2020.

\bibitem[Bonawitz et~al.(2016)Bonawitz, Ivanov, Kreuter, Marcedone, McMahan,
  Patel, Ramage, Segal, and Seth]{bonawitz2016practical}
K.~A. Bonawitz, Vladimir Ivanov, Ben Kreuter, Antonio Marcedone, H.~Brendan
  McMahan, Sarvar Patel, Daniel Ramage, Aaron Segal, and Karn Seth.
\newblock Practical secure aggregation for federated learning on user-held
  data.
\newblock In \emph{NIPS Workshop on Private Multi-Party Machine Learning},
  2016.

\bibitem[Carlini et~al.(2020)Carlini, Deng, Garg, Jha, Mahloujifar, Mahmoody,
  Song, Thakurta, and Tramer]{carlini2020attack}
Nicholas Carlini, Samuel Deng, Sanjam Garg, Somesh Jha, Saeed Mahloujifar,
  Mohammad Mahmoody, Shuang Song, Abhradeep Thakurta, and Florian Tramer.
\newblock An attack on instahide: Is private learning possible with instance
  encoding?
\newblock In \emph{IEEE Symposium on Security and Privacy}, 2020.

\bibitem[Chen et~al.(2021)Chen, Li, Song, and Zhuo]{chen2020instahide}
Sitan Chen, Xiaoxiao Li, Zhao Song, and Danyang Zhuo.
\newblock On instahide, phase retrieval, and sparse matrix factorization.
\newblock In \emph{ICLR}, 2021.

\bibitem[Cohen et~al.(2019)Cohen, Lee, and Song]{cls19}
Michael~B Cohen, Yin~Tat Lee, and Zhao Song.
\newblock Solving linear programs in the current matrix multiplication time.
\newblock In \emph{Proceedings of the 51st Annual ACM Symposium on Theory of
  Computing (STOC)}, 2019.

\bibitem[Commission(2018)]{gdpr}
European Commission.
\newblock 2018 reform of eu data protection rules.
\newblock https://gdpr-info.eu/, 2018.

\bibitem[Deng et~al.(2009)Deng, Dong, Socher, Li, Li, and
  Fei-Fei]{deng2009imagenet}
Jia Deng, Wei Dong, Richard Socher, Li-Jia Li, Kai Li, and Li~Fei-Fei.
\newblock Imagenet: A large-scale hierarchical image database.
\newblock In \emph{CVPR}, 2009.

\bibitem[Deng(2012)]{deng2012mnist}
Li~Deng.
\newblock The mnist database of handwritten digit images for machine learning
  research.
\newblock \emph{IEEE Signal Processing Magazine}, 29\penalty0 (6):\penalty0
  141--142, 2012.

\bibitem[Dwork(2009)]{d09}
Cynthia Dwork.
\newblock The differential privacy frontier.
\newblock In \emph{Theory of Cryptography Conference (TCC)}, pages 496--502,
  2009.

\bibitem[Dwork and Roth(2014)]{dr14}
Cynthia Dwork and Aaron Roth.
\newblock The algorithmic foundations of differential privacy.
\newblock \emph{Foundations and Trends in Theoretical Computer Science},
  9\penalty0 (3--4):\penalty0 211--407, 2014.

\bibitem[Geiping et~al.(2020)Geiping, Bauermeister, Dr{\"o}ge, and
  Moeller]{geiping2020inverting}
Jonas Geiping, Hartmut Bauermeister, Hannah Dr{\"o}ge, and Michael Moeller.
\newblock Inverting gradients--how easy is it to break privacy in federated
  learning?
\newblock In \emph{NeurIPS}, 2020.

\bibitem[He et~al.(2016)He, Zhang, Ren, and Sun]{he2015resnet}
Kaiming He, Xiangyu Zhang, Shaoqing Ren, and Jian Sun.
\newblock Deep residual learning for image recognition.
\newblock In \emph{CVPR}, 2016.

\bibitem[Huang et~al.(2020)Huang, Song, Li, and Arora]{huang2020instahide}
Yangsibo Huang, Zhao Song, Kai Li, and Sanjeev Arora.
\newblock Instahide: Instance-hiding schemes for private distributed learning.
\newblock In \emph{ICML}, 2020.

\bibitem[Ioffe and Szegedy(2015)]{ioffe2015batch}
Sergey Ioffe and Christian Szegedy.
\newblock Batch normalization: Accelerating deep network training by reducing
  internal covariate shift.
\newblock In \emph{ICML}, 2015.

\bibitem[Kairouz et~al.(2021)Kairouz, McMahan, Avent, Bellet, Bennis, Bhagoji,
  Bonawitz, Charles, Cormode, Cummings, et~al.]{kairouz2019advances}
Peter Kairouz, H~Brendan McMahan, Brendan Avent, Aur{\'e}lien Bellet, Mehdi
  Bennis, Arjun~Nitin Bhagoji, Keith Bonawitz, Zachary Charles, Graham Cormode,
  Rachel Cummings, et~al.
\newblock Advances and open problems in federated learning.
\newblock \emph{Foundations and Trends in Machine Learning}, 14\penalty0
  (1–2):\penalty0 1--210, 2021.

\bibitem[Kingma and Ba(2015)]{kingma2014adam}
Diederik~P Kingma and Jimmy Ba.
\newblock Adam: A method for stochastic optimization.
\newblock In \emph{ICLR}, 2015.

\bibitem[Krizhevsky et~al.(2009)]{cifar10}
Alex Krizhevsky et~al.
\newblock Learning multiple layers of features from tiny images.
\newblock 2009.

\bibitem[Lamb et~al.(2019)Lamb, Verma, Kannala, and
  Bengio]{lamb2019interpolated}
Alex Lamb, Vikas Verma, Juho Kannala, and Yoshua Bengio.
\newblock Interpolated adversarial training: Achieving robust neural networks
  without sacrificing too much accuracy.
\newblock In \emph{Proceedings of the 12th ACM Workshop on Artificial
  Intelligence and Security}, pages 95--103, 2019.

\bibitem[Legislature(2018)]{ccpa}
California~State Legislature.
\newblock California consumer privacy act.
\newblock \url{https://oag.ca.gov/privacy/ccpa}, 2018.

\bibitem[Li et~al.(2021)Li, Jiang, Zhang, Kamp, and Dou]{li2021fedbn}
Xiaoxiao Li, Meirui Jiang, Xiaofei Zhang, Michael Kamp, and Qi~Dou.
\newblock Fedbn: Federated learning on non-iid features via local batch
  normalization.
\newblock In \emph{ICLR}, 2021.

\bibitem[McMahan et~al.(2016)McMahan, Moore, Ramage, Hampson,
  et~al.]{mcmahan2016communication}
H~Brendan McMahan, Eider Moore, Daniel Ramage, Seth Hampson, et~al.
\newblock Communication-efficient learning of deep networks from decentralized
  data.
\newblock In \emph{Artificial Intelligence and Statistics (AISTATS)}, pages
  1273--1282, 2016.

\bibitem[Pang et~al.(2020)Pang, Xu, and Zhu]{pang2019mixup}
Tianyu Pang, Kun Xu, and Jun Zhu.
\newblock Mixup inference: Better exploiting mixup to defend adversarial
  attacks.
\newblock In \emph{ICLR}, 2020.

\bibitem[Papernot et~al.(2020)Papernot, Chien, Song, Thakurta, and
  Erlingsson]{papernot2020making}
Nicolas Papernot, Steve Chien, Shuang Song, Abhradeep Thakurta, and Ulfar
  Erlingsson.
\newblock Making the shoe fit: Architectures, initializations, and tuning for
  learning with privacy, 2020.
\newblock URL \url{https://openreview.net/forum?id=rJg851rYwH}.

\bibitem[Phong et~al.(2017)Phong, Aono, Hayashi, Wang, and Moriai]{phong_17}
Le~Trieu Phong, Yoshinori Aono, Takuya Hayashi, Lihua Wang, and Shiho Moriai.
\newblock Privacy-preserving deep learning: Revisited and enhanced.
\newblock In \emph{ICATIS}, pages 100--110, 2017.

\bibitem[Phong et~al.(2018)Phong, Aono, Hayashi, Wang, and Moriai]{phong18}
Le~Trieu Phong, Yoshinori Aono, Takuya Hayashi, Lihua Wang, and Shiho Moriai.
\newblock Privacy-preserving deep learning via additively homomorphic
  encryption.
\newblock \emph{IEEE Transactions on Information Forensics and Security}, 2018.

\bibitem[Shamir(1979)]{shamir1979share}
Adi Shamir.
\newblock How to share a secret.
\newblock \emph{Communications of the ACM}, 22\penalty0 (11):\penalty0
  612--613, 1979.

\bibitem[Tram{\`e}r and Boneh(2021)]{tramer2020differentially}
Florian Tram{\`e}r and Dan Boneh.
\newblock Differentially private learning needs better features (or much more
  data).
\newblock In \emph{ICLR}, 2021.

\bibitem[Wei et~al.(2020)Wei, Liu, Loper, Chow, Gursoy, Truex, and
  Wu]{wei2020framework}
Wenqi Wei, Ling Liu, Margaret Loper, Ka-Ho Chow, Mehmet~Emre Gursoy, Stacey
  Truex, and Yanzhao Wu.
\newblock A framework for evaluating gradient leakage attacks in federated
  learning.
\newblock \emph{arXiv preprint arXiv:2004.10397}, 2020.

\bibitem[Yin et~al.(2021)Yin, Mallya, Vahdat, Alvarez, Kautz, and
  Molchanov]{yin2021see}
Hongxu Yin, Arun Mallya, Arash Vahdat, Jose~M Alvarez, Jan Kautz, and Pavlo
  Molchanov.
\newblock See through gradients: Image batch recovery via gradinversion.
\newblock \emph{arXiv preprint arXiv:2104.07586}, 2021.

\bibitem[Zhang et~al.(2018{\natexlab{a}})Zhang, Cisse, Dauphin, and
  Lopez-Paz]{zhang2017mixup}
Hongyi Zhang, Moustapha Cisse, Yann~N Dauphin, and David Lopez-Paz.
\newblock Mixup: Beyond empirical risk minimization.
\newblock In \emph{ICLR}, 2018{\natexlab{a}}.

\bibitem[Zhang et~al.(2019)Zhang, Dauphin, and Ma]{zhang2019fixup}
Hongyi Zhang, Yann~N Dauphin, and Tengyu Ma.
\newblock Fixup initialization: Residual learning without normalization.
\newblock In \emph{ICLR}, 2019.

\bibitem[Zhang et~al.(2018{\natexlab{b}})Zhang, Isola, Efros, Shechtman, and
  Wang]{zhang2018unreasonable}
Richard Zhang, Phillip Isola, Alexei~A Efros, Eli Shechtman, and Oliver Wang.
\newblock The unreasonable effectiveness of deep features as a perceptual
  metric.
\newblock In \emph{CVPR}, 2018{\natexlab{b}}.

\bibitem[Zhao et~al.(2020)Zhao, Mopuri, and Bilen]{zhao2020idlg}
Bo~Zhao, Konda~Reddy Mopuri, and Hakan Bilen.
\newblock idlg: Improved deep leakage from gradients.
\newblock \emph{arXiv preprint arXiv:2001.02610}, 2020.

\bibitem[Zhu and Blaschko(2021)]{zhu2020r}
Junyi Zhu and Matthew Blaschko.
\newblock R-gap: Recursive gradient attack on privacy.
\newblock In \emph{ICLR}, 2021.

\bibitem[Zhu et~al.(2019)Zhu, Liu, and Han]{zhu2020deep}
Ligeng Zhu, Zhijian Liu, and Song Han.
\newblock Deep leakage from gradients.
\newblock In \emph{NeurIPS}, 2019.

\end{thebibliography}
